\documentstyle[multicol,aps,prl,eqsecnum,floats,epsf]{revtex}
\begin{document}
\renewcommand{\thesection}{\arabic{section}}

\draft
\title{
%{\tenrm\hfill PRE submitted}\\
NORMAL AND ANOMALOUS SCALING OF THE FOURTH-ORDER CORRELATION FUNCTION OF A
RANDOMLY ADVECTED PASSIVE SCALAR}
\author{M. Chertkov$^a$, G. Falkovich$^a$, I. Kolokolov$^b$
and V. Lebedev$^{a,c}$}
\address{
$^a$ Department of Physics of Complex Systems, Weizmann Inst. of Science,
Rehovot 76100, Israel \\ $^b$ Univ. di Milano, INFN,
via Celoria 16, Milano 20133 Italia \\ and Budker Inst. of
Nuclear Physics, Novosibirsk 630090, Russia \\ $^c$ Landau
Inst. for Theor. Physics, Moscow, Kosygina 2, 117940 Russia}
%\date{\today}
\maketitle
\begin{abstract}
For a delta-correlated velocity field, simultaneous correlation functions
of a passive scalar satisfy closed equations. We analyze the equation for the
four-point function. To describe a solution completely, one has to solve the
matching problems at the scale of the source and at the diffusion scale.
We solve both the matching problems and thus
find the dependence of the four-point correlation function on the diffusion
and pumping scale for large space dimensionality $d$.
It is shown that anomalous scaling
appears in the first order of $1/d$ perturbation theory.
Anomalous dimensions are found analytically both for the scalar field and for
it's derivatives, in particular, for the dissipation field.
\end{abstract}


\begin{multicols}{2}

\section*{Introduction}
\label{sec:intro}

It is already a commonplace to talk about an anomalous scaling of the
high-order correlation functions in developed turbulence.
By this, the deviation of the scaling exponents from their ``naive''
values taken from dimensional estimate or perturbation theory is usually
meant. Another meaning ascribed to that term is related to the cases
where the exponent of the $2n$-th correlation function is not $n$ times
the exponent of the second one so that the degree of non-Gaussianity depends
on scale. The experimental evidence in favor of anomalous
scaling of a scalar field advected by turbulence exists for some time
\cite{84AHGA,90MS,91Sre,91HY} while the attempts of the consistent derivation
(starting from the equations of fluid mechanics) of the
correlation functions of the order higher than two started only recently
\cite{93Maj,94Kra,95KYC,95CFKL,Proc,95GK}.
The problem of a passive scalar advection, being of physical importance by
itself,
may serve also as a starting point in studying anomalous scaling in turbulence
\cite{94Kra,94LPF}. At a first step, we found \cite{94CFKL} the whole set of
the simultaneous correlation functions for the Batchelor-Kraichnan problem
of a scalar advected by a large-scale random velocity field
\cite{59Bat-a,74Kra-a}.
It has been shown that, whatever be the (finite)
temporal correlations of the velocity field,
all correlation functions of the scalar are integer powers of a logarithm for
all
the distances in the convective interval of scales and no anomalous
scaling thus appears at the leading terms.
The present paper is an account of the next step: we consider multi-scale
velocity field with power spectrum. Following Kraichnan \cite{68Kra-a},
we restrict ourselves by the simplest possible temporal behavior assuming both
velocity field and scalar source to be white in time. That leads to a
substantial simplification of the analytical description since any simultaneous
correlation function of a scalar satisfies closed  linear differential equation
of
the second order (see \cite{94Kra} and below).

In an isotropic turbulence, n-point correlation function depends on $n(n-1)/2$
distances for a dimensionality of space $d>n-2$.
For the pair correlation function, the respective ordinary differential
equation could be readily solved for any distance between the points ---
see \cite{68Kra-a} and (\ref{a12}) below. The solution is expressed via
the flux of a squared scalar $P_2$ and molecular and eddy diffusivity, the
scaling exponent $\zeta_2$ in the inertial interval being fixed by the
condition of the flux constancy.  If the scaling exponent of the $2n$-th
correlation function is $n\zeta_2$, that is called a normal scaling. An
anomalous
scaling would mean that the true answer has additional factors like
$(l/r)^\Delta$
where $\Delta$ is anomalous dimension, $r$ is the distance between points and
$l$ is some length parameter. One may imagine three
reasons for anomalous scaling according to the three lengths that may be
relevant:
i) $l$ is the pumping scale $L$ so that the anomalous scaling appears due to
infrared non-locality, ii) $l$ is the diffusion scale $r_d$ so that the
anomalous
scaling appears due to ultraviolet non-locality \cite{94LL,94LPF}
and iii) anomalous scaling appears
due to the existence of the high-order integrals of motion \cite{94Fal-a} so
that
the length parameter appears from the ratio of the different fluxes.
First nontrivial object that may reveal anomalous scaling is the
fourth-order correlation function. One may be interested in two-point objects
like $\langle(\theta_1-\theta_2)^4\rangle$ or pair correlation
function $\langle\epsilon_1\epsilon_2\rangle$ of the dissipation
field $\epsilon(t,{\bf r})=\kappa[\nabla\theta(t,{\bf r})]^2$
yet they do not satisfy any
closed equation. To find such two-point-fourth-order objects, one should solve
the complete equation for the four-point function and then fuse some points.
Considering $d>2$, one has to deal with the space of six variables which
makes the direct solution at arbitrary parameters quite difficult. Gawedski and
Kupiainen \cite{95GK} recently developed the perturbation theory that describes
the effect of the weak advection on the fourth-order correlation function
for a non-cascade diffusion-like regime and found the respective anomalous
exponents.
Our main target here is the description of the cascade in the
convective interval of scales, we solve the equation for the four-point
correlation
function assuming the space dimensionality $d$ to be large. Non-gaussian part
of the correlation function is small in parameter $1/d$ which makes it possible
to develop a regular perturbation theory.
In the present paper, we  obtain the scaling exponents analytically in the
first
order in $1/d$ (assuming $1/d$ to be the smallest parameter, in particular,
$1/d\ll\zeta_2$) and show that the main contribution into the general
fourth-order
correlation function  has $L$-related anomalous scaling with
\begin{equation}
\Delta={4(2-\zeta_2)\over d}\ .\label{Delta}\end{equation}
Note that $\Delta$ turns into zero when $\zeta_2\rightarrow2$
which contradicts Kraichnan's closure \cite{94Kra,95KYC} and is in
a qualitative agreement with Gawedski and
Kupiainen result \cite{95GK} even though they considered a different limit.

For the correlation functions between the
points separated by some distances from the convective interval and by others
from
a diffusion interval of scales, we show that the main term does not depend on
$r_d$.
We also show that  the subleading terms in the case of some
distances being much smaller than others necessarily have another anomalous
exponents with respect to small distances. Such subleading terms determine,
in particular, correlations of the spatial derivatives of the scalar field,
which thus have an $r_d$-related anomalous scaling (in addition to a general
$L$-related scaling).
Note that the results on $r_d$-dependence are obtained without $1/d$-expansion
so they are exact at any dimension.
For example, the dissipation field has an $r_d$-related
anomalous
scaling with the exponent that is equal to the scaling exponent of
the eddy diffusivity so that the irreducible correlator
$\langle\langle\epsilon_1\epsilon_2\rangle\rangle\propto (L/r_{12})^\Delta
r_d^0$
does not depend on the diffusion scale $r_d$ when the distance $r_{12}$ is
in the convective interval.
Another consequence of our results is the statement that
$\langle\epsilon^2\rangle/\langle
\epsilon\rangle^2$ tends to infinity with Peclet number increasing.
Contrary,
the one-point statistics of $\theta$ (say, the flatness
$\langle\theta^4\rangle/
\langle\theta^2\rangle^2$) is independent of the Peclet number $Pe=L/r_d$
at the limit of large $Pe$.

Besides, we find the nontrivial $r_d$-related
anomalous scaling describing the traceless tensor
%% FOLLOWING LINE CANNOT BE BROKEN BEFORE 80 CHAR
$\nabla_\alpha\theta\nabla_\beta\theta-d^{-1}\delta_{\alpha\beta}(\nabla\theta)^2$.
We also argue that high-order integrals of motion may influence scaling if the
pumping is non-Gaussian and that the different kinds of
$L$-related anomalous scaling may be
the case starting from the sixth correlation function.

The anomalous $L$-related
scaling of the fourth-order correlation function of the scalar field
confronted with
the diagrammatic analysis that shows no infrared divergences at any finite
order of a Wyld diagram technique
\cite{94LPF} evidently means that such an anomalous scaling is a
non-perturbative phenomenon.

The structure of the article is as follows. We formulate the problem and find
the
pair correlation function in Section~\ref{sec:formul}, the second section is
devoted to $L$-dependence while the third section to $r_d$ dependence
of the fourth-order correlation functions. The correlation functions of
the scalar derivatives are also considered in the third section.
Section \ref{sec:PG} describes
possible generalizations and Conclusion summarizes the results.

\section{Formulation of the problem}
\label{sec:formul}

We consider the advection of a passive scalar
field $\theta(t,{\bf r})$ by an incompressible turbulent flow.
The advection is governed by the following equation
\begin{eqnarray}
&&(\partial_t-\hat{P})\theta=\phi,
\label{a1a}\\
&&\hat{P}(t)=-u^{\alpha}\nabla^{\alpha}+\kappa \triangle,\qquad
\nabla^\alpha u^\alpha=0,\label{a1}
\end{eqnarray}
where both convective and diffusive terms are included in
the operator $\hat{P}(t)$, $\triangle$ designates Laplacian here.
The external velocity ${\bf u}(t,{\bf r})$ and the external
source $\phi(t,{\bf r})$ are random functions of $t$ and ${\bf r}$.
We regard the statistics of the velocity and of the source to be independent.
Therefore all correlation functions of $\theta$ are to be treated as
averages over both statistics. Averaging over pumping will be designated by
over-line and averaging over velocity will be designated by angular brackets.

\subsection{Basic Relations}

We assume that the source $\phi$ is $\delta$-correlated in time
and spatially correlated on a scale $L$. The latter means e.g. that the pair
correlation function of the source
\begin{equation}
\overline{\phi(t_1,{\bf r}_1)\phi(t_2,{\bf r}_2)}=
\delta(t_1-t_2)\chi(r_{12}),
\label{a3}
\end{equation}
as a function of the argument $r_{12}\equiv|{\bf r}_1-{\bf r}_2|$
decays on the scale $L$. The value $\chi(0)=P_2$
is the production rate of $\theta^2$.
Since the pumping is Gaussian then high-order
correlation functions are determined by $\chi(r)$. For example,
\begin{eqnarray}
&&
\overline{\phi(t_1,{\bf r}_1)\phi(t_2,{\bf r}_2)
\phi(t_3,{\bf r}_3)\phi(t_4,{\bf r}_4)}\nonumber\\&&=
\chi( r_{12})\chi(r_{34})\delta(t_1-t_2)\delta(t_3-t_4)
\nonumber\\&&
+ \chi( r_{13})\chi( r_{24})\delta(t_1-t_3)\delta(t_2-t_4)
\nonumber\\&&
+ \chi( r_{14})\chi(r_{23})\delta(t_1-t_4)\delta(t_2-t_3).
\label{xi4} \end{eqnarray}

A formal solution of (\ref{a1a}) is
\begin{eqnarray}
&& \theta(t,{\bf r})=\int_{-\infty}^t dt' \hat{U}(t,t')\phi(t',{\bf r}),
\label{a2a}\\
&& \hat{U}(t,t')={\sl T}\exp\biggl(\int_{t'}^{t}
\hat{P}(\tau)d\tau\biggr)\equiv
\label{a2} \\ &&
\sum_{n=0}^{\infty}\int_{t'}^td\tau_1\int_{t'}^{\tau_1}d\tau_2\dots
\int^{\tau_{n-1}}_{t'}d\tau_n\hat{P}(\tau_1)\dots\hat{P}(\tau_n).
\nonumber
\end{eqnarray}
Since $\theta$ and $\phi$ are related linearly then the pair product
of the passive scalar field averaged over the statistics of pumping is
expressed
via the pair correlation function of the pumping
\begin{eqnarray}&&
\overline{\theta(t,{\bf r}_1)\theta(t,{\bf r}_2)}
=\int\limits_{-\infty}^t d\tau \hat{U}_{1}(t,\tau)
\hat{U}_{2}(t,\tau)
\chi(r_{12})\nonumber\\&&
=\int\limits_{-\infty}^t d\tau {\sl T}\exp\biggl(\int_{\tau}^{t}
\bigl[\hat{P}_{1}(\tau)+
\hat{P}_{2}(\tau)\bigr]d\tau\biggr)\chi(r_{12}).\label{a4}
\end{eqnarray}
Here and below $\hat U_i,\hat P_i$ designate operators acting on variables
${\bf r}_i$. Time-ordered exponents in the r.h.s. of (\ref{a4}) commute with
 each other (because they have different space arguments), that allows to
rewrite in (\ref{a4}) their product as a single ${\sl T}\exp$.

The next
step is to average ${\sl T}\exp$ in (\ref{a4}) with respect to the statistics
of the velocity field ${\bf u}$. Following Kraichnan, we consider
the case of a velocity delta-correlated in time but {\em multi-scale} in space.
Velocity statistics is completely determined by the pair correlation function
\begin{eqnarray} &&
\langle  u^\alpha(t_1,{\bf r}_1) u^\beta(t_2,{\bf r}_2)\rangle=
\delta(t_1-t_2)V^{\alpha\beta}\,,
\nonumber \\ &&
V^{\alpha\beta}=V_0
\delta^{\alpha\beta}-{\cal K}^{\alpha\beta}({\bf r}_{12}),\quad
{\cal K}^{\alpha\beta}(0)=0\ .
\label{a5a}\end{eqnarray}
Here the so-called eddy diffusivity is as follows
\begin{eqnarray}
{\cal K}^{\alpha\beta}=\frac{D}{r^\gamma}
(r^2\delta^{\alpha\beta}-r^\alpha r^\beta)+\frac{D(d-1)}{2-\gamma}
\delta^{\alpha\beta} r^{2-\gamma}\,.\label{K}
\end{eqnarray}
where  $0<\gamma<2$ is supposed and $\langle\dots\rangle$ stands for an
average over the velocity field statistics, isotropy  being assumed.
The representation (\ref{K}) is valid for the scales less than velocity
infrared cut-off $L_u$, which is supposed to be the largest scale of the
problem.
The marginal ``logarithmic'' $\gamma=0$ and ``diffusive'' $\gamma=2$ regimes
require a special care.

\subsection{Simultaneous pair correlator}

Let us calculate the simultaneous pair correlation function of the
passive scalar
\begin{equation}
f(r_{12})=\langle\overline{\theta({\bf r}_1)\theta({\bf r}_2)}\rangle\,.
\label{a6} \end{equation}
Since we assume the statistics of
the velocity field to be delta-correlated and Gaussian, then
the $n$-th order correlator of the velocity field [which appears in
the expansion of the ${\sl T}\exp$ from the integrand of (\ref{a4})]
can be found explicitly by the Wick theorem which prescribes to
reduce any average to a product of pair correlation functions.
Those terms are summed up into a usual operator exponent
\begin{eqnarray}
&&\biggl\langle {\sl T}\exp\biggl(\int\limits_{\tau}^{t} [\hat{P}_{1}(\tau)+
\hat{P}_{2}(\tau)]d\tau\biggr)\biggr\rangle =
e^{(t-\tau)\hat{\cal L}_{12}},
\label{a7a}\\&&
\hat{\cal L}_{ij}\equiv V^{\alpha\beta}({\bf r}_{ij})
\nabla^\alpha_{i}\nabla^\beta_{j}+
\Bigl(\frac{V_0}{2}+\kappa\Bigr)(\triangle_{i}+\triangle_{j})\,,
\label{a7} \end{eqnarray}
where ${\bf r}_{ij}={\bf r}_{i}-{\bf r}_{j}$ and $\triangle_{i}=\nabla_i^2$.
We find from (\ref{a4})
\begin{equation}
f(r_{12})=\int_{-\infty}^t d\tau
\exp\biggl((t-\tau)\hat{\cal L}_{12}\biggr)
\chi(r_{12}) \,.
\label{a7b} \end{equation}
Integrating the r.h.s. of (\ref{a7b}) with respect to time
($\hat{\cal L}^{-1}_{ij}$ is a well defined integral operator) we get
$\hat{\cal L}_{12}f(r_{12})=-\chi(r_{12})$ which is an ordinary second-order
differential equation
\begin{mathletters} \label{a9} \begin{eqnarray} &&
\hat{\cal L}^{(p)}f=-\chi \,, \qquad {\rm where}
\qquad \partial_r\equiv \partial/\partial r
\label{a9a} \\ &&
\hat{\cal L}^{(p)}=\frac{(d-1)D}
{2-\gamma}r^{1-d}\partial_r \bigl(r^{2-\gamma}
+r_d^{2-\gamma}\bigr)r^{d-1}\partial_r \,,
\label{a9b} \end{eqnarray} \end{mathletters}

\noindent
previously derived by Kraichnan \cite{68Kra-a}. In (\ref{a9b})
the diffusion scale $r_d$ was introduced
\begin{equation}
r^{2-\gamma}_d=\frac{2\kappa(2-\gamma)}{D(d-1)}.
\label{a10} \end{equation}
Equation (\ref{a9}) is integrated explicitly.
The solution is completely determined by two physical boundary conditions:
zero at $r=\infty$ and absence of a singularity at $r=0$. We consider
the pumping correlation
function close to a step function $\chi(r)=P_2$ at $r<L$ and zero otherwise:
\begin{eqnarray} &&
f(r)=P_2\int_r^\infty g(r')dr'\,,
\label{a11} \\
\underline{\mbox{at}\quad r>L}\qquad &&
g(r)=\frac{L^d}{2\kappa d}
\frac{r^{1-d}}{1+(r/r_d)^{2-\gamma}}\,,
\nonumber \\
\underline{\mbox{at}\quad r<L}\qquad &&
g(r)=\frac{r}{2\kappa d}\frac{1}{1+(r/r_d)^{2-\gamma}}\, .
\nonumber \end{eqnarray}
For a general $\gamma$, the pair correlation function is expressed via
a transcendental function
\begin{eqnarray}
f(r)={P_2r_d^\gamma\over d(d-1)}\biggl\{{\pi\over\sin[2\pi/(2-\gamma)]}
\nonumber \\
-\Phi\bigl[-\bigl({r/ r_d}\bigr)^{2-\gamma},1,{2/(2-\gamma})
\bigr]\biggr\} \,,
\label{a111} \end{eqnarray}
-- see \cite{PBM1}, p.29 and \cite{Bat1} p.27.
In the Richardson-Kolmogorov case, $d=3,\gamma=2/3$ and one gets
\begin{equation}
f(0)-f(r)={P_2\over 3}\Bigl[r^{2/3}-r_d^{2/3}{\rm arctg}\,(r/r_d)^{2/3}\Bigr]\
{}.
\label{a112}\end{equation}
Both (\ref{a111}) and (\ref{a112}) are valid at any $r\leq L$  for the
step-like
pumping. At $r\ll L$, those formulas describe the pair correlation function
for the case of an arbitrary pumping $\chi$ with the correlation scale $L$.
One can see that the transition from convective to diffusive interval is
described by a universal function.

If the Peclet number $Pe=L/r_d$ is large, the pumping and diffusion scales
separate three intervals of scales with a different
scaling behavior:
\begin{mathletters} \label{a12}
\begin{eqnarray} &&
\underline{ r\ll r_d\ll L}\qquad f(r)\approx P_2\biggl(
\frac{L^\gamma(2-\gamma)}{\gamma(d-\gamma)(d-1)D}-
\frac{r^2}{4\kappa d}\biggr),
\nonumber \\ && \label{a12a} \\ &&
\underline{ r_d\ll r< L}\qquad
f(r)\approx P_2\frac{2-\gamma}{\gamma(d-1)D}
\biggl(\frac{L^\gamma}{d-\gamma}-\frac{r^\gamma}{d}\biggr),
\nonumber\\&&\label{a12b} \\ &&
\underline{ r_d\ll L< r}\qquad f(r)\approx P_2
\frac{L^d(2-\gamma){r^{\gamma-d}}}{d(d-1)(d-\gamma)D}.
\label{a12c} \end{eqnarray}
\end{mathletters}

\noindent
Those expressions fully determine the behavior of the pair
correlation function.

\subsection{Different-time pair correlator}

The dynamical analog of (\ref{a4}) looks as follows $(t_2>t_1)$
\begin{eqnarray} &&
\overline{\theta(t_1,{\bf r}_1)\theta(t_2,{\bf r}_2)}
\nonumber\\&&=\hat{U}_{2}(t_2,t_1)\int_{-\infty}^t d\tau
\hat{U}_{1}(t_1,\tau)\hat{U}_{2}(t_1,\tau) \chi(r_{12})
\nonumber \\ &&=
\hat{U}_{2}(t_2,t_1)
\overline{\theta(t_1,{\bf r}_1)\theta(t_1,{\bf r}_2)} \,.
\label{D1} \end{eqnarray}
Since the  velocity field is delta-correlated, then one may reduce the average
of the product from the r.h.s. of (\ref{D1}) to the product of the averages
\begin{eqnarray} &&
\langle \overline{\theta(t_1,{\bf r}_1)\theta(t_2,{\bf r}_2)}\rangle
=\langle\hat{U}_{2}(t_2,t_1)\rangle f(r_{12})\nonumber \\ &&=
\exp\biggl(\mid t_2-t_1\mid(V_0+\kappa)\triangle\biggr)f(r_{12}).\nonumber
\end{eqnarray}
Let us stress that if $r_d\ll L$ then $V_0\gg \kappa$. The point is that
$V_0$ can be estimated as $V_0\sim D L_u^{2-\gamma}$ where $L_u$
is the scale of the largest vortices. Since $L_u>L$ is assumed then
$V_0\gg \kappa$. That means that the time dependence of correlation
functions of the passive scalar $\theta$ is determined by the velocity of
the largest vortices and is therefore fast. Note that in the comoving
reference frame this dynamics is much more slow \cite{94LPF}.

\section{Four-point correlation function of the passive scalar}
\label{sec:four}

We begin the analysis of the fourth-order correlation function $F_{1234}$.
It will be based upon the equation (\ref{b6}), previously derived
independently by
Kraichnan \cite{91Kra-b}, Sinai and Yakhot \cite{SY88} and Shraiman and Siggia
\cite{SS94}. It is a second-order partial differential
equation in the space of $4d$ variables. Isotropy allows one to diminish
the number of variables to five at $d=2$ and to six at $d>2$, which is
still too many to hope to find $F$ explicitly at all possible distances
$r_{ij}$ ranging from zero to infinity. Our aim is modest: we are looking
for the scaling exponents only. First, we shall find the overall scaling
exponent $\zeta_4$ which describes how $\Gamma$ (which is the
irreducible part of $F_{1234}$) scales if all the distances $r_{ij}$ are
multiplied by the same factor. This is the subject of this section where
we employ $1/d$ perturbation theory assuming space dimensionality to be large.
That will also allow us to find the anomalous exponents that describe
the $L$-dependence of the correlation function.
Second, we shall consider the case with one or two distances being much
less than other ones and find the scaling exponents with respect to small
and large distances separately, this could be done at an arbitrary $d$.
That will allow us to fuse some points
and to find the $r_d$-dependence of the cumulants of second powers of
spatial derivatives. This is the subject of the next section.

\subsection{The equation for the simultaneous four-point correlation function}

Using the technique developed in the preceding section we derive the
equation for the simultaneous four-point correlation function
$F_{1234}$. Let us form first a four-point object averaged over
the pumping only by analogy with (\ref{a4})
\begin{eqnarray} &&
\overline{\theta(t,{\bf r}_1)\theta(t,{\bf r}_2)
\theta(t,{\bf r}_3)\theta(t,{\bf r}_4)}=
\nonumber \\ &&
=\int_{-\infty}^td\tau_1\int_{-\infty}^td\tau_2\int_{-\infty}^td\tau_3
\int_{-\infty}^td\tau_4 \hat{U}_{1}(t,\tau_1)
\hat{U}_{2}(t,\tau_2)
\nonumber \\ &&
\hat{U}_{3}(t,\tau_3)\hat{U}_{4}(t,\tau_4)
\overline{\phi(\tau_1,{\bf r}_1)\phi(\tau_2,{\bf r}_2)\phi(\tau_3,{\bf r}_3)
\phi(\tau_4,{\bf r}_4)}.
\label{b1} \end{eqnarray}
Due to (\ref{xi4}) the object separates onto three parts
$Q_{12;34}+Q_{13;24}+Q_{14;23}$, where
\begin{eqnarray} &&
Q_{ij;kl}\!\!=\!\!\int\limits_{-\infty}^t\!\!\!\int\!\! d\tau d\tilde{\tau}
\hat{U}_{i}(t,\tau)\hat{U}_{j}(t,\tau)
\hat{U}_{k}(t,\tilde{\tau})\hat{U}_{l}(t,\tilde{\tau})
\chi(r_{ij})\chi(r_{kl})\nonumber \\ &&
=\!\!\int\limits_{-\infty}^t\!\! d\tau {\sl
T}\exp\biggl[\int\limits_{\tau}^td\tau'
(\hat{P}_{i}\!+\!\hat{P}_{j}\!+\!\hat{P}_{k}\!+\!
\hat{P}_{l})\biggr]\!
\int\limits_{-\infty}^{\tau}\!d\tilde{\tau}
\biggl[{\sl T}\exp\biggl(\int\limits_{\tilde{\tau}}^\tau\!\!
d\tilde{\tau}'\nonumber \\ &&
\times(\hat{P}_{i}\!+\!\hat{P}_{j})\biggr)\!
+\!{\sl T}\exp\biggl(\int\limits_{\tilde{\tau}}^\tau\!\! d\tilde{\tau}'
(\hat{P}_{k}\!+\!\hat{P}_{l})\biggr)
\biggr]\chi(r_{ij})\chi(r_{kl}).
\label{b2} \end{eqnarray}

\noindent
The commutativity of $\hat{P}_i$ and $\hat{P}_j$ at $i\neq j$
has been taken into account at (\ref{b2}).

For the delta-correlated velocity field, one may
average the time-ordered exponents in the r.h.s. of (\ref{b2}) explicitly,
first, expanding them into series and, second,
performing the Gaussian decomposition of the velocity correlators
in the same manner as it was done for the pair correlator of the passive
scalar in (\ref{a7a}). That results in the following expressions
\begin{eqnarray} &&
\langle{Q_{ij;kl}}\rangle
=\int_{-\infty}^t d\tau e^{(t-\tau)\hat{\cal L}}\times
\nonumber\\ &&
\int_{-\infty}^{\tau}d\tilde{\tau} \biggl(e^{(\tau-\tilde{\tau})\hat{\cal
L}_{12}}+
e^{(\tau-\tilde{\tau})\hat{\cal L}_{34}}\biggr)
\chi(r_{ij})\chi(r_{kl})\,,
\label{b3a} \end{eqnarray}
where $\hat{\cal L}_{ij}$ was defined in (\ref{a7}) and $\hat{\cal L}$ has
the following form:
\begin{equation}
\hat{\cal L}\equiv \frac{1}{2}\sum_{i,j}
V^{\alpha\beta}({\bf r}_{ij})
\nabla^\alpha_{i}\nabla^\beta_{j}+
\kappa\sum\triangle_{i}\,,
\label{b4} \end{equation}
where $\nabla^\alpha_i\equiv \partial/\partial r_{i}^\alpha$.
The integration over $\tilde\tau$ in (\ref{b3a}) gives
\begin{eqnarray}
-(\hat{\cal L}_{ij}^{-1}\!+\!\hat{\cal L}_{kl}^{-1})
\chi(r_{ij})\chi(r_{kl})\!=\!
% \nonumber\\&&
f(r_{ij})\chi(r_{kl})\!+\!\chi(r_{ij})f(r_{kl})
\nonumber \end{eqnarray}
where $\hat{\cal L}_{ij}f(r_{ij})=-\chi(r_{ij})$ was used.
Inserting the last expressions into (\ref{b3a})
and collecting all the terms $\langle{Q_{ij;kl}}\rangle$
we find  for the full four-point object
$F_{1234}\equiv\langle\overline{\theta({\bf r}_1)\theta({\bf r}_2)
\theta({\bf r}_3)\theta({\bf r}_4})\rangle$:
\begin{eqnarray}
F_{1234}=\int\limits_{-\infty}^{t} d\tau\,
\exp\bigl[(t-\tau)\hat{\cal L}\bigr]
\sum_{\{ij\}\neq\{kl\}}f(r_{ij})\chi(r_{kl}),
\nonumber \end{eqnarray}
where the notation $\{ij\}$ stands for a pair of non-coinciding ($i\neq j$)
site indices. That leads to the following equation
\begin{eqnarray}
&& -\hat{\cal L} F_{1234}=
f(r_{12})\chi(r_{34})+f(r_{34})\chi(r_{12})+
f(r_{13})\chi(r_{24})\nonumber\\&&+f(r_{24})\chi(r_{13})+
f(r_{14})\chi(r_{23})+f(r_{23})\chi(r_{14}),
\label{b6} \end{eqnarray}
with the r.h.s. expressed in terms of the functions $f(r_{ij})$ already
found: see (\ref{a11},\ref{a12}). Note that at a confluence of the points
${\bf r}_i\rightarrow{\bf r}_j$ the r.h.s. of  (\ref{b6}) tends to a constant.
Using (\ref{a9}) one can obtain an equation for the irreducible four-point
correlator $\Gamma_{1234}=\langle\langle
\overline{\theta({\bf r}_1)\theta({\bf r}_2)
\theta({\bf r}_3)\theta({\bf r}_4})\rangle\rangle\equiv
F_{1234}-f_{12}f_{34}-f_{13}f_{24}-f_{14}f_{23}$:
\begin{mathletters} \label{b7}
\begin{eqnarray} &&
-\hat{\cal L}\Gamma_{1234}=
\Phi_{12;34}+\Phi_{13;24}+\Phi_{14;23},
\label{b7a}\\ &&
\Phi_{ij;kl}=\Bigl[{\cal K}^{\alpha\beta}({\bf r}_{il})
-{\cal K}^{\alpha\beta}({\bf r}_{ik})
+{\cal K}^{\alpha\beta}({\bf r}_{jk})-
{\cal K}^{\alpha\beta}({\bf r}_{jl})\Bigr]
\nonumber\\ &&
\quad\times\nabla^\alpha_{ij}
\nabla^\beta_{kl}f({\bf r}_{ij})f({\bf r}_{kl})\,,
\label{b7b} \end{eqnarray}
\end{mathletters}

\noindent
where $\nabla^\alpha_{ij}\equiv \partial/\partial r_{ij}^\alpha$.

The operator $\hat{\cal L}$ (\ref{b4}) is negatively defined. To prove this
we represent $\hat{\cal L}=\kappa\hat{\cal L}_{dif}+\hat{\cal L}_c$ where
the convective part could be written as $\hat{\cal L}_c=
\bigl\langle\bigl[\sum{\bf u}({\bf r}_i)\nabla_i\bigr]^2\bigr\rangle_u$.
Here we designate $\langle\ldots\rangle_v$ as an average
over the velocity field ${\bf u}({\bf r})$ random in space -- from
the viewpoint of initial ${\bf u}({\bf r},t)$, the averaging is over the
instant
configuration. Incompressibility guarantees that  $\hat{\cal L}_c$ is Hermitian
and
negatively defined as well as $\hat{\cal L}_{dif}$. Therefore, $\hat{\cal L}$
has
a continuous non-positive spectrum  and the density of states regular at zero.
The last statement follows from the inequality
$\| \kappa\hat{\cal L}_{dif}+\hat{\cal L}_c \|
\geq\| \kappa\hat{\cal L}_{dif}\|$ and the absence of singularity in
the density of states for $\hat{\cal L}_{dif}$. We can thus conclude that
the equation $-\hat{\cal L}\Gamma=\Phi$ is well defined for $\Gamma$ and $\Phi$
that do not grow at infinity.

The operator $\hat{\cal L}$ is scale invariant if all the distances $r_{ij}$
are
either much larger than $r_d$ or much smaller. The right-hand side is scale
invariant if all the distances are either larger or smaller than $L$.
We thus could divide our space of $r_{ij}$
into three domains where the scale invariance of $\Gamma(r_{ij})$
is to be expected. It is natural to ask now a simple question:
what prevents us from making the statement that the scaling exponent
of the solution is equal to the difference between the exponents of the
right-hand side and of the operator $\hat{\cal L}$? Of course, the
solutions with such ``naive'' scalings exist in all three regions
(we call them forced solutions $F_{forc}$). The problem is that to match those
particular solutions at pumping and diffusion scale, it may be necessary to
include
into the full solution the zero modes of the operator:
\begin{equation}
F=F_{forc}+{\cal Z}\ .\label{zemo}
\end{equation}
Here the zero mode ${\cal Z}$ may have a scaling different from that of
$F_{forc}$.
To avoid misunderstanding,
note that since the operator $\hat{\cal L}$ is non-positive then it
cannot have a global zero mode that satisfies boundary conditions.
The parts of the solution with an anomalous scaling may nevertheless
be considered as zero modes within separate domains.
At $r_{ij}\ll L$, for instance, one should worry about zero modes that
appear due to matching conditions at $r_{ij}\simeq L$, such modes are
allowed to grow with $r_{ij}$. At an oversimplified
level, this is what happens when we obtain the constant
$L$-dependent term in the expressions (\ref{a12a},\ref{a12b})
for the pair correlation function. In a multidimensional
space, operators may have zero modes much more complicated than a constant and,
indeed, the operator $\hat{\cal L}$ does have an infinite number of zero modes.

We cannot yet find the zero modes and solve the matching
problem analytically at arbitrary $d$. Fortunately, for $d>2$ (generally, for
$d>n-2$ where $n$ is the number of points in the correlation function) we found
the representation of $\hat{\cal L}$ that allows one to represent
$\Gamma(r_{ij})$ (with all distances of the same order or the main term if
some distances are smaller than others) as a power
series with respect to some numerical parameter ($1/d$ in
the convective interval). The coefficients in the series are
functions of $r_{ij}$ all having the same scaling exponents and logarithms
that appear from expanding anomalous exponents in powers of $1/d$. By
analyzing those functions at the next subsection, we establish
the overall scaling properties of $\Gamma(r_{ij})$ in the convective and
diffusion intervals. The possibility to sum the logarithms into
a power functions is provided by an explicit scale invariance of the
equation that determines the zero mode in the convective interval.
The consideration of the subleading terms at the case of
some distances being small in comparison with other ones requires a special
approach that will be developed in Section \ref{sec:twopa}.

\subsection{Representation of Tetrahedron's Lengths}
\label{subsec:perturb}

To establish overall scaling properties of the irreducible fourth-order
correlator $\Gamma$ it will be convenient for us to use a special
representation of the equations (\ref{b6},\ref{b7}). In an isotropic
case, the four-point correlation
function $\Gamma$ is a function of six distances $r_{ij}$
between the points, $d>2$ being assumed. The case $d=2$ where there are only
five independent variables deserves special consideration. In the variables
$r_{ij}=|{\bf r}_{ij}|$, the operator $\hat{\cal L}$ is a sum of two parts,
$\hat{\cal L}=\hat{\cal L}_0+\hat{\cal L}_1$:
\begin{eqnarray} &&
\hat{\cal L}_0=\frac{D(d-1)}
{2-\gamma}\sum_{i>j} r^{1-d}_{ij}\partial_{r_{ij}} \bigl(r^{2-\gamma}_{ij}
+r_d^{2-\gamma}\bigr)r^{d-1}_{ij}\partial_{r_{ij}}
\nonumber \\ &&
\hat{\cal L}_1=-\frac{D(d-1)}{2(2-\gamma)} \sum
(r_{in}^2-r_{ij}^2-r_{jn}^2)
\frac{r_{ij}^{1-\gamma}}{r_{jn}}
\frac{\partial^2}{\partial r_{ij}\partial r_{jn}}
\nonumber \\ &&
-\frac{D}{4}\sum
\frac{1}{r_{ij}^\gamma r_{im}r_{jn}}\biggl(
\frac{d+1-\gamma}{2-\gamma}r_{ij}^2
(r_{in}^2+r_{jm}^2-r_{ij}^2-r_{mn}^2)
\nonumber \\ &&
+\frac{1}{2}(r_{ij}^2+r_{im}^2-r_{jm}^2)
(r_{ij}^2+r_{jn}^2-r_{in}^2)\biggr)
\frac{\partial^2}{\partial r_{im}\partial r_{jn}}
\nonumber \\ &&
+\kappa\sum \frac{r_{ij}^2+r_{im}^2-r_{mj}^2}{2r_{ij} r_{im}}
\frac{\partial^2}{\partial r_{ij}\partial r_{im}} \,.
\label{h2} \end{eqnarray}
Here the summation is performed over subscripts satisfying
the conditions $i\neq j$ and $m\neq i,j$ , $n\neq i,j$.

What is nice about that representation is that if we omit $\hat{\cal L}_1$
from $\hat{\cal L}$ then the reducible part
$F^{(0)}_{1234}=f_{12}f_{34}+f_{13}f_{24}+f_{14}f_{23}$
of the fourth-order correlator appears to be a solution of (\ref{b6}). It is
\begin{eqnarray}&&
F^{(0)}=\sum f(r_{ij})f(r_{kl})=(\mbox{in the conv. int.})\nonumber\\&&=
\frac{3(2-\gamma)^2L^{2\gamma}}{\gamma^2(d-1)^2(d-\gamma)^2}\!+\!
\frac{(2-\gamma)^2}{\gamma^2(d-1)^2d^2}\sum \!{\cal
Z}_{ij,kl}\!+\!F_{forc},\nonumber\\&&
{\cal Z}_{ij,kl}=r_{ij}^\gamma r_{kl}^\gamma-\frac{d}{2(d+\gamma)}
\bigl(r_{ij}^{2\gamma}+r_{kl}^{2\gamma}\bigr),\label{zm3}\\&&
F_{forc}\!=\!\frac{(2-\gamma)^2\sum
r_{ij}^{2\gamma}}{2\gamma^2(d+\gamma)(d-1)^2d}\!
-\!\frac{(2-\gamma)^2 L^\gamma\sum
r_{ij}^{\gamma}}{\gamma^2(d-1)^2d(d-\gamma)}\ .
\label{F0a}
\end{eqnarray}
which is the zero approximation of (\ref{zemo}).
Here ${\cal Z}_{ij,kl}$ is the zero mode of $\hat{\cal L}_0$ while
that is remarkable that a direct check shows $F_{forc}$ to be a partial
solution
in the convective interval
of the full equation $-\hat{\cal L} F_{forc}=f(r_{12})\chi(r_{34})+\cdots$.
It means, particularly, that if we develop
the iteration procedure in $\hat{\cal L}_0^{-1}\hat{\cal L}_1$  then
all the terms appeared at higher
steps will enter zero modes of the full operator
$\hat{\cal L}$ but not the forced term found.
Another remarkable feature of the zero step is an absence among
the terms three-point zero modes of $\hat{\cal L}_0$, like
${\cal Z}_{ij,ik}$. They will appear on the next step only.

The basic fact is that $\hat{\cal L}_0\propto d^2$ while
$\hat{\cal L}_1\propto d$ as $d\rightarrow\infty$. Assuming $1/d$ to be a
formal
small parameter, we shall implement the iteration procedure with respect to
$\hat{\cal L}_1$
\begin{equation}
\hat{\cal L}_0\Gamma^{(1)}=
-\hat{\cal L}_1F^{(0)} \,, \quad
\hat{\cal L}_0 \Gamma^{(n)}=
- \hat{\cal L}_1\Gamma^{(n-1)} \,.
\label{h3} \end{equation}
This procedure leads to a representation of $\Gamma$ as a series over the
powers of $1/d$:
\begin{equation}
\Gamma=\sum_{n=1}^\infty(-\hat{\cal L}_0^{-1}
\hat{\cal L}_1)^n F^{(0)} \,.
\label{h4} \end{equation}
which is actually the series for the nongaussian part of the zero mode.

One may note that the four-point correlation function is defined not in the
whole
six-dimensional space of $r_{ij}$ but rather in the physical subspace
restricted by
triangle inequalities $r_{ij}+r_{jk}\geq r_{ik}$. That makes no additional
difficulties since the solution we shall find satisfies all boundary
conditions in the physical subspace. In addition, the
answer is expressed in terms of powers of
$\hat{\cal L}_0^{-1}\hat{\cal L}_1$ which do not have singularities at
the boundary of the subspace.

At each step of the iteration procedure we should solve the equation of
the (\ref{h3}) type. Such equation is much easier to analyze
than e.g. (\ref{b7a}). The point is that the operator $\hat{\cal L}_0$
is a sum of the six sub-operators $\hat{\cal L}^{(p)}$ from (\ref{a9b})
each depending on the single variable $r_{ij}$ only.
This particular form of $\hat{\cal L}_0$ enables us to analyze the zero modes
and establish the necessary properties of its resolvent. The solution of the
equation (\ref{h3}) is expressed via the corresponding integral kernel as
follows
\begin{equation}
\Gamma^{(n)}(\vec r)=
\int dt\prod_{i>j}\int dr^\prime_{ij}
{\cal R}(t;\vec r,\vec r\,^\prime)
\hat{\cal L}_1 \Gamma^{(n-1)}(\vec r\,^\prime) \,,
\label{h5} \end{equation}
where $\vec r$ designates the set of six variables $r_{ij}$ and
the integration is performed over time $t$ and six separations
$r^\prime_{ij}$. The resolvent ${\cal R}$ can be represented as a product
\begin{equation}
{\cal R}(t;\vec r,\vec r\,^\prime)=
\prod_{i>j} {\cal R}^{(p)}(t; r_{ij}, r^\prime_{ij}) \,,
\label{h6} \end{equation}
where the function ${\cal R}^{(p)}(t;r,r^\prime)$ satisfies the equation
\begin{equation}
(\partial_t-\hat{\cal L}_r^{(p)})
{\cal R}^{(p)}(t;r,r^\prime)=\delta(t)\delta(r-r^\prime) \,,
\label{r1} \end{equation}
with the condition ${\cal R}^{(p)}(t;r,r^\prime)=0$ at $t<0$.
Note the essential property of ${\cal R}$
simplifying the subsequent analysis: it is independent of
the pumping scale $L$. That means that there is only one characteristic
length in ${\cal R}$: the diffusion scale introduced by (\ref{a10}).

Now we are going to establish the properties of the resolvent
${\cal R}^{(p)}(t;r,r^\prime)$. Let us remind that there are two
terms in the operator $\hat{\cal L}^{(p)}$ (\ref{a9b}) which are of
the same order at the diffusion scale $r_d$. Solving (\ref{r1}) at
the diffusion and convection intervals, we get for the resolvent
\begin{mathletters} \label{r3}
\begin{eqnarray} &&
\underline{\mbox{at}\quad r,r^\prime,\sqrt{t\kappa}\ll r_d}
\label{r3a} \\ &&
{\cal R}^{(p)}\approx\frac{1}{4 t\kappa}
\exp\biggl(-\frac{r^2+r^{\prime 2}}{8t\kappa}\biggr)
I_{d/2-1}\biggl(\frac{rr^\prime}{4t\kappa}\biggr)
r^{1-d/2}r^{\prime d/2},
\nonumber \\ &&
\underline{\mbox{at}\quad r,r^\prime,(Dt)^{1/\gamma}\gg r_d}
\nonumber \\ &&
{\cal R}^{(p)}\approx\frac{(2-\gamma)}{\gamma(d-1)tD}
r^{(\gamma-d)/2}r^{\prime(d+\gamma-2)/2}\times
\label{r3b} \\ &&
\exp\biggl(-\frac{(r^\gamma+r^{\prime^\gamma})
(2-\gamma)}{\gamma^2(d-1)tD}\biggr)I_{d/\gamma-1}
\biggl(\frac{2(2-\gamma)(rr^\prime)^{\gamma/2}}
{\gamma^2(d-1)tD}\biggr) \,.
\nonumber \end{eqnarray}
\end{mathletters}

\noindent
We shall need also the asymptotics in the
mixed limit $r, r_d\ll r^\prime, (Dt)^{1/\gamma}$, for $r$ and $r^\prime$
lying in the different intervals (diffusion and convective ones respectively).
This asymptotics has to match with one following from (\ref{r3b})
in the sub-interval $r_d\ll r\ll r^\prime,(Dt)^{1/\gamma}$.
It gives an idea: to improve the expansion of (\ref{r3b}) with respect to
$r/r^\prime$ introducing a function $y(r)$:
\begin{mathletters} \label{r4-5}
\begin{eqnarray} &&
\underline{\mbox{at}\quad r, r_d\ll r^\prime, (Dt)^{1/\gamma}}
\nonumber \\ &&
{\cal R}^{(p)}\approx\left[\frac{(2-\gamma)}{\gamma^2(d-1)tD}\right]^{d/\gamma}
{\gamma r^{\prime(d-1)}\over \Gamma(d/\gamma)}\exp\biggl(
-\frac{r^{\prime\gamma}(2-\gamma)}{\gamma^2(d-1)tD}\biggr)
\nonumber\\&&\times
\biggl\{1+\biggl[-1+\frac{2(2-\gamma)r^{\prime\gamma}}{d(d-1)\gamma t}\biggr]
\frac{2(2-\gamma)y(r)}{(d-1)\gamma t}\biggr\}\,.
\label{r4} \end{eqnarray}
In the limit $r_d\ll r$, the function $y(r)$ should pass to $r^\gamma$ as it
follows from (\ref{r3b}). Substituting (\ref{r4}) into (\ref{r1}) and solving
the resulting equation for $y$ one gets an expression with one unknown
parameter:
\begin{equation}
y(r)=y(0)+(d-1)D\int_0^r\frac{xdx}
{D(d-1)x^{2-\gamma}+2\kappa(2-\gamma)} \,.
\label{r5} \end{equation}
\end{mathletters}

\noindent
Generally, the integration of the second order differential equation
produces two parameters, but here  one of them has been already fixed by the
condition of the finiteness of $y(r)$ at $r\to 0$.
The only scale which can determine the dimensional parameter $y(0)$ is $r_d$.
It gives the following estimate $y(0)\sim r_d^\gamma$ and, thus, closes our
analysis of the mixed asymptotics of the resolvent at
$r, r_d\ll r^\prime, (Dt)^{1/\gamma}$.

\subsection{Overall Scaling of the Fourth-Order Correlator}
\label{subsec:gen}

The analysis of the series (\ref{h4}) enables us to establish the
scaling behavior of the irreducible part of the fourth-order correlator
$\Gamma$ in different regions of scales. The crucial point
is that all terms of the series (\ref{h4}) have the same scaling
behavior as $\Gamma^{(1)}$ up to some logarithmic functions that are
our main concern. Indeed, we should worry if some logarithms could appear at
each
step of the iteration procedure (\ref{h3}), summation of powers of logarithms
can produce anomalous exponents changing the index of the whole sum in
comparison with the first term of (\ref{h4}).

We shall analyze in this Section only logarithms $\ln(L/r_{ij})$.
Besides, the logarithms of the ratios of separations $\ln(r_{ij}/r_{kl})$
may arise at any step of the iteration procedure (\ref{h3}) which could
change the behavior at small ratios. This happens only in subleading
terms where the coefficient at logarithms are proportional to a power of
the small ratio. One may worry, nevertheless, if
the logarithms could be summed into the
large exponent which compensates the small factors. That means that the
case of small ratios needs a separate non-perturbative analysis which
is done in Section \ref{sec:twopa}. The results of the analysis show
that after re-summation the subleading terms remain the subleading ones.
Therefore, the logarithms of the ratios can be neglected at the general
investigation. Below in this section, we will speak about
the {\it overall scaling} of the zero modes that is about their
exponents in terms of the ratio $L/r$.

First, we show that the zero modes of $\hat{\cal L}_0$ cannot contain
logarithms
$\ln(L/r_{ij})$.
It can be proved using
an ``angular'' representation in $r_{ij}$-space introduced as
\begin{equation}
r_{12}^{\gamma/2}=R^{\gamma/2} n_1\,,
\quad r_{13}^{\gamma/2}=R^{\gamma/2} n_2 \,, \quad \dots \,,
\label{gen2} \end{equation}
where $R$ is a ``radial'' variable and $\vec n$ is a unit $6$-component vector.
In terms of those variables the operator $\hat{\cal L}_0$ at $R\gg r_d$
is written as
$$\hat{\cal L}_0=\frac{(d-1)D}{(2-\gamma)R^\gamma}
\left(R^2\partial_R^2 +(6d+1-\gamma)R\partial_R
+\hat\Upsilon \right) \,,$$
where $\hat\Upsilon$ is an angular operator:
\begin{eqnarray} &&
\hat\Upsilon =\frac{\gamma(1-\gamma+d)}{2}
\sum_{m=1}^6\left( (n_m^{-1}-6n_m)
\frac{\partial}{\partial n_m}\right) +
\nonumber \\ &&
(\gamma/2)^2\sum_{m=1}^6 n_m^{2/\gamma-1}
\frac{\partial^\perp}{\partial n_m}
n_m^{1-2/\gamma}\frac{\partial^\perp}{\partial n_m} \,,
\label{gen4} \\ &&
\frac{\partial^\perp}{\partial n_m}
\equiv \frac{\partial}{\partial n_m}
-\sum_{l=1}^6 n_m n_l \frac{\partial}{\partial n_l} \,.
\nonumber \end{eqnarray}
The consequence of (\ref{gen4}) is a possibility to look for
the zero modes in the following form:
\begin{eqnarray} &&
{\cal Z}_n = R^{a_n} Y_n (\vec n) \,,
\label{gen5} \\ &&
\hat\Upsilon Y_n (\vec n)=\lambda_n Y_n (\vec n) \,,
\label{gen6} \\ &&
a_n^2+ (6d-\gamma) a_n +\lambda_n =0 \,.
\label{gen7} \end{eqnarray}
The roots of the equation (\ref{gen7}) determine exponents in the expression
(\ref{gen5}) for the zero modes which appear to be power functions
[like (\ref{zm3})]. Dangerous logarithms would occur
if the roots of (\ref{gen7}) coincide. That
corresponds to the values $a_{n,cr}=-(6d-\gamma)/2 $,
$\lambda_n=(6d-\gamma)^2/4$, which is impossible since all
eigenvalues $\lambda_n$ of the operator $\hat\Upsilon$ are negative.
The property follows from the fact that all eigenvalues of
$\hat{\cal L}_0$ are negative and $\hat{\cal L}_0$ is reduced to
$\hat\Upsilon$ for $R$-independent functions. Nondegeneracy of the
eigenvalues guarantees the absence of the logarithms $\ln(L/r_{ij})$
in the zero modes of $\hat{\cal L}_0$. Note that the coincidence of $a_n$
for two different $\lambda_n$ (i.e. the existence of two
zero modes with the same scaling yet different angular structures)
apparently does not lead to the appearance of logarithms as well.

To establish the exponent of the overall scaling,
let us analyze the integral (\ref{h5}). First,
we show that at any step of the iteration the integral
converges at $r^\prime\gg L$ so that this region gives negligible
contribution if we consider separations $r_{ij}\lesssim L$.
To do that, one should find the behavior of $\Gamma^{(n)}$ at large scales.
The decay of $\Phi$ from (\ref{b7b}) is different for different direction
in the six-dimensional space $\{r_{ij}\}$. By analyzing the
most slowly decaying case we get $\Phi\sim R^{-\gamma}$ where
$R$ is determined by (\ref{gen2}). Now we can analyze (\ref{h5}) and
using the resolvent which is determined for all scales one can directly
show that the region $R^\prime\gg L$ gives a negligible contribution.
Then the analysis of $\Gamma^{(1)}$ in the region $R \gg L$ shows that
$\Gamma^{(1)}$ decays like $\Phi$. That means that the above scheme is
reproduced at any step of the iteration procedure.

Consider the case where all $r_{ij}$ are in the convective
interval. Then we can use (\ref{r3b}) for ${\cal R}^{(p)}$.
The contribution to $\Gamma^{(1)}$ associated with $r^\prime<r$ is
$L$-independent.
One can check using (\ref{r3b}) and (\ref{a12b}) that the integral in
(\ref{h5})
converges at small $r^\prime$ already in the convective interval.
That means that the contribution associated
with $r^\prime$ from the convective interval
is also $r_d$-independent and it is thus given by the
simple dimensional estimate: $P_2^2/D^2$ multiplied by a scale invariant
function of $r_{ij}$ with the exponent $2\gamma$. That contribution
could be neglected in comparison with the terms proportional to
positive powers of $L$ which will be found below. Using the expressions
analogous to (\ref{r4-5}) for small $r^\prime$ one can also check
that the contribution to the integral due to the diffusion region is
finite and negligible.
The main contribution to $\Gamma^{(n)}$ at the integration
over $r^\prime_{ij}$ in (\ref{h5}) is given by the region
$r^\prime_{ij}\sim L$, the characteristic time being estimated as
$t\sim L^\gamma/D$ . It follows then from (\ref{r3b})  that for
$r\ll L$ the corresponding expression for  ${\cal R}^{(p)}$ is
a regular expansion in $r^\gamma$, what means that the contribution
to $\Gamma^{(1)}$ associated with the considered region
$r^\prime_{ij}\sim L$ is also a regular expansion in $r_{ij}^\gamma/L^\gamma$.
Only the first terms, proportional to positive powers of $L$,
are important, the subsequent terms proportional to negative powers of
$L$ can be neglected in the convective interval.

The zero term in $r^\gamma$-expansion of (\ref{h5}) is a
constant which can be estimated as $P_2^2L^{2\gamma}/D^2$.
The first term (that would be proportional to $L^\gamma r^\gamma$)
is actually equal to zero.
That could be understood as follows.
As it is seen from (\ref{a12b}, \ref{h2})
in the convective interval, the r.h.s of  (\ref{h3}) does not depend on $L$:
(\ref{a12b}) contains only an $L$-dependent constant, these constants
has to be killed in (\ref{h5}) since the operator (\ref{h2}) contains
only cross derivatives. That $L$-independence of the r.h.s. means that an
$L$-dependent term in $\Gamma^{(1)}$ should be treated as a zero mode (up to a
possible logarithmic factor)
of the operator $\hat{\cal L}_0$ (in the convective interval only!).
Zero modes with the scaling $r^\gamma$ exist (e.g.
$r_{12}^\gamma-r_{34}^\gamma$) but one cannot construct such mode
symmetric in permutations of the points $1,2,3,4$. Thus the presence of
such zero modes would violate the intrinsic ${\bf r}_i\leftrightarrow {\bf
r}_j$
symmetry of the problem.

The next term behaves $\propto r^{2\gamma}$.
If $\Gamma^{(1)}$ contains logarithmic factors multiplied by
$r^{2\gamma}$ then the normal scaling is
renormalized into an anomalous one.
Let us consider the general expression for $\Gamma^{(1)}$
\begin{equation}
\Gamma^{(1)}(\vec{r})=\int^\infty dt \int d\vec{r}'{\cal R}(t;\vec{r},\vec{r}')
\hat{\cal L}_1 F^{(0)}(\vec{r}')\ ,
\label{1}
\end{equation}
expand it in the series in $r_{ij}^\gamma$ and calculate (for instance)
the terms proportional to $r_{12}^\gamma r_{34}^\gamma,
r_{12}^\gamma r_{13}^\gamma$ and $r_{12}^{2\gamma}$ respectively:
\end{multicols}
\begin{eqnarray}&&
\Delta \Gamma^{(1)}_{12,34}=
r_{12}^\gamma r_{34}^\gamma
\int \frac{dt}{t} {\cal A}_1(tD/L^\gamma),\quad
\Delta \Gamma^{(1)}_{12,13}=
r_{12}^\gamma r_{13}^\gamma
\int \frac{dt}{t} {\cal A}_2(tD/L^\gamma),\quad
\Delta \Gamma^{(1)}_{12,12}=
r_{12}^{2\gamma}
\int \frac{dt}{t} {\cal A}_3(tD/L^\gamma),\label{2a}\\&&
{\cal A}_1(t/L^\gamma)=\int^L d r'_{12} \tilde{\cal R}^{(p)}(t;0,r'_{12})
\int^L d r'_{34} \tilde{\cal R}^{(p)}(t;0,r'_{34})
\int^L d r'_{13} {\cal R}^{(p)}(t;0,r'_{13})
\int^L d r'_{14} {\cal R}^{(p)}(t;0,r'_{14})\times\nonumber\\&&\times
\int^L d r'_{23} {\cal R}^{(p)}(t;0,r'_{23})
\int^L d r'_{24} {\cal R}^{(p)}(t;0,r'_{24}) \hat{\cal
L}_1(\vec{r}')F^{(0)}(\vec{r}'),\label{2ba}\\&&
{\cal A}_2(t/L^\gamma)=\int^L d r'_{12} \tilde{\cal R}^{(p)}(t;0,r'_{12})
\int^L d r'_{13} \tilde{\cal R}^{(p)}(t;0,r'_{13})
\int^L d r'_{34} {\cal R}^{(p)}(t;0,r'_{34})
\int^L d r'_{14} {\cal R}^{(p)}(t;0,r'_{14})\times\nonumber\\&&\times
\int^L d r'_{23} {\cal R}^{(p)}(t;0,r'_{23})
\int^L d r'_{24} {\cal R}^{(p)}(t;0,r'_{24}) \hat{\cal
L}_1(\vec{r}')F^{(0)}(\vec{r}'),\label{2bb}\\&&
{\cal A}_3(t/L^\gamma)=\int^L d r'_{12} \tilde{\tilde{\cal
R}}^{(p)}(t;0,r'_{12})
\int^L d r'_{34} {\cal R}^{(p)}(t;0,r'_{34})
\int^L d r'_{13} {\cal R}^{(p)}(t;0,r'_{13})
\int^L d r'_{14} {\cal R}^{(p)}(t;0,r'_{14})\times\nonumber\\&&\times
\int^L d r'_{23} {\cal R}^{(p)}(t;0,r'_{23})
\int^L d r'_{24} {\cal R}^{(p)}(t;0,r'_{24}) \hat{\cal
L}_1(\vec{r}')F^{(0)}(\vec{r}'),\label{2bc}\\&&
{\cal R}^{(p)}(t;0,x)=\biggl[\frac{2-\gamma}{\gamma^2(d-1)
tD}\biggr]^{d/\gamma}
\frac{\gamma}{\Gamma[d/\gamma]}
\exp\biggl[-x^\gamma\frac{2-\gamma}{\gamma^2(d-1)tD}\biggr]
x^{d-1},\label{2ca}\\&&
\tilde{\cal R}^{(p)}(t;0,x)={\cal
R}^{(p)}(t;0,x)\frac{(2-\gamma)}{(d-1)\gamma^2}
\biggl[-1+\frac{(2-\gamma)x^\gamma}{d(d-1)\gamma tD}\biggr],
\label{2cb}\\&&
\tilde{\tilde{\cal R}}^{(p)}(t;0,x)=
{\cal R}^{(p)}(t;0,x)\frac{(2-\gamma)^2}{2(d-1)^2\gamma^4}
\biggl[1-\frac{2(2-\gamma)x^\gamma}{d(d-1)\gamma tD}+
\frac{(2-\gamma)^2x^{2\gamma}}{d(d-1)^2(d+\gamma)\gamma^2 t^2D^2}\biggr],
\label{2cc}
\end{eqnarray}
\begin{multicols}{2}
While integrating over time at  $tD\ll L^\gamma$ one can expand
the functions ${\cal A}_i(tD/L^\gamma)$ in a regular series in $tD/L^\gamma$:
\begin{equation}
{\cal A}_i(tD/L^\gamma)=C^{(i)}_0+C^{(i)}_1 tD/L^\gamma+\cdots.
\label{3}
\end{equation}
The crucial point is the presence of the zero term in the
expansion. Were $C^{(i)}_0$ be equal to zero we would immediately figure out
the part of the well known zero-mode
from the divergent integral over $t$ ($\sim C^{(i)}_1$) in (\ref{2a}),
the respective
term in (\ref{2a}) being $L$-independent.
As we show below, $C^{(i)}_0\neq 0$ so that one gets
the logarithm $\ln[L/r]$
in the respective ($\sim C^{(i)}_0$) contribution to $\Delta\Gamma_i^{(1)}$.
This term stems from the forced solution of the equation
$\hat{\cal L}_0\Gamma^{(1)}=-\hat{\cal L}_1F^{(0)}$ (it is proved already that
there are no logarithms in zero modes of $\hat{\cal L}_0$). From the viewpoint
of the initial equation (2.5), those logarithms stem from the zero modes of the
full operator ${\cal L}$ --- see (\ref{F0a}) and below.

Let us calculate $C^{(i)}_0$ directly. First of all one finds
from (\ref{h2})
\begin{eqnarray}&&
-\hat{\cal L}_1F^{(0)}\!=\!-\biggl[\frac{2-\gamma}{\gamma(d-1)d}\biggr]^2
\hat{\cal L}_1\bigl[r_{12}^\gamma r_{34}^\gamma+
r_{13}^\gamma r_{24}^\gamma+
r_{14}^\gamma r_{23}^\gamma\bigr]\nonumber\\&&=
\frac{(2-\gamma)^2}{8(d-1)^2d^2}\sum
\biggl(
2\frac{d+1-\gamma}{2-\gamma}r_{ij}^2
(r_{in}^2+r_{jm}^2-r_{ij}^2-r_{mn}^2)\nonumber\\&&
+(r_{ij}^2\!+\!r_{im}^2\!-\!r_{jm}^2)
(r_{ij}^2\!+\!r_{jn}^2\!-\!r_{in}^2)\biggr)
r_{ij}^{-\gamma}r_{im}^{\gamma-2}r_{jn}^{\gamma-2},
\label{L1F}
\end{eqnarray}
where the summation is performed over all the sets of different subscripts
$\{i,j,m,n\}=\{1,2,3,4\}$. $F^{(0)}(\vec{r})$ and respectively the integrands
in
(\ref{2ca}-\ref{2cc}) are sums of a huge number of very simple terms that
have been calculated by Mathematica.
We do not present here the cumbersome expressions obtained formally at
arbitrary
$\gamma$ and $d$ since the physical meaning could be ascribed to them
only when ${\cal L}_1/{\cal L}_0$-expansion is a small-parameter expansion.
The below formulas show that there are two cases where this is so:
$C^{(i)}_0$ go to zero at $\gamma\to 2$ and
at $d\to\infty$:
\begin{eqnarray}&&
\mbox{at}\quad d\to\infty\quad C^{(1)}_0\to
\frac{4(2-\gamma)^3}{\gamma^3d^5},
\label{dinf1}\\&&
\mbox{at}\quad d\to\infty\quad C^{(2)}_0\to
-\frac{2(2-\gamma)^3}{\gamma^3d^5},
\label{dinf2}\\&&
\mbox{at}\quad d\to\infty\quad C^{(3)}_0\to\frac{2(2-\gamma)^3}{\gamma^3 d^5},
\label{dinf3}\\&&
\mbox{at}\quad \gamma\to 2^-\quad C^{(1)}_0\to-\frac{(5d-2-d^2)(2-\gamma)^3}
{2(d-2)(d-1)^3d^3},
\label{gm21}\\&&
\mbox{at}\quad \gamma\to 2^-\quad C^{(2)}_0\to
-\frac{(2+d^2-4d)(2-\gamma)^3}{4(d-2)(d-1)^3d^3},
\label{gm22}\\&&
\mbox{at}\quad \gamma\to 2^-\quad
C^{(3)}_0\to+\frac{(2-\gamma)^3}{4d^2(d-1)(d-2)}.
\label{gm23}
\end{eqnarray}
It is natural that our $\hat{\cal L}_0^{-1}\hat{\cal L}_1$-expansion gives
small correction in the limit $d\rightarrow\infty$ while we should
admit that the same behavior in the limit $\gamma\rightarrow2$ is surprising
for us. That means that $\hat{\cal L}_0^{-1}\hat{\cal L}_1$-expansion
might be meaningful also
when $d$ is arbitrary while $2-\gamma$ is small. To establish that,
one should check whether nonlogarithmic terms in $\Gamma^{(1)}$
are proportional to $(2-\gamma)^3$ as well, which is beyond the scope
of the present approach. A nontrivial
task to develop a consistent
$(2-\gamma)$-expansion will be the subject of further publications.
As $\gamma\rightarrow2$ velocity spectrum diverges at small scales,
one should introduce ultraviolet cut-off $r_\eta$ in
(\ref{K}) so that $r^{2-\gamma}/(2-\gamma)\rightarrow\ln(r/r_\eta)$.
The scalar correlation functions contains logarithms already at $\gamma=2$
(in the
paper \cite{95GK}, the limit $\gamma\rightarrow2,D/(2-\gamma)\rightarrow
$const has been considered where such logarithms are absent).

Contrary, the perturbation theory in $1/d$ which is the main subject of this
Section is regular and uniform with respect to distances $r_{ij}$.
Note that we assume $1/d$ to be the smallest
parameter in the problem, in particular, $1/d\ll\gamma$
so that our results do
not describe the limit of small $\gamma$
 which will be the subject of further
publications.

We have found all the logarithmic terms entering the fourth-order correlator
on the first step of the iteration procedure:
\begin{eqnarray}
&&F_{12,34}=F^{(0)}_{12,34}+
\gamma\biggl\{C_0^{(1)}\sum r_{ij}^\gamma r_{kl}^\gamma\ln[L/r]\nonumber\\&&
+
C_0^{(2)}\sum r_{ij}^\gamma r_{ik}^\gamma\ln[L/r]+
C_0^{(3)}\sum r_{ij}^{2\gamma}\ln[L/r]\biggr\}\nonumber\\&&
+\mbox{$L$-independent or non-logarithmic terms}+\cdots,
\label{FF}
\end{eqnarray}
where the summations are performed over all the possible combinations
supposing all the indexes $i,j,k,l$ to be different.
The expression in front of the logarithms in (\ref{FF})
is a sum of the well-known zero
modes ($\sim r^{2\gamma}$), that is guaranteed by
the identity that could be directly checked
\begin{equation}
C_0^{(3)}+\frac{d}{2(d+\gamma)}\bigl(C_0^{(1)}+4C_0^{(2)}\bigr)=0.
\label{ident}
\end{equation}
It agrees with the absence of any logarithms among zero modes
of $\hat{\cal L}_0$.

We have thus found the first
(in $1/d$) term of the expansion of the
zero-mode part of the solution of the full operator
\begin{eqnarray}&&
{\cal Z}=\frac{(2-\gamma)^2}{\gamma^2d^4}
\sum Z_{ij,kl}+\gamma
\biggl(C_0^{(1)}\sum{\cal Z}_{ij,kl}\ln[L/r]+\nonumber\\&&
C_0^{(2)} \sum{\cal Z}_{ij,ik}\ln[L/r]
\biggr)+\cdots. \label{zr}
\end{eqnarray}
Note that there are two types ($Z_{ij,kl}$ and $Z_{ij,ik}$) of zero modes
of $\hat{\cal L}_0$ possessing scaling $2\gamma$, only one of them
($Z_{ij,kl}$) is present on the zero step of the iteration procedure.
It is clear that (\ref{zr}) cannot be described in terms of a single anomalous
exponents $\Delta(d)$ because the expression in the last brackets does not
coincide with $\sum Z_{ij,kl}$.
We thus come to the conclusion that ${\cal Z}$ should consist of at least two
terms with different scaling. On another language it means that
one gets a degenerate case. Indeed, The full operator $\hat{\cal L}$,
the bare operator $\hat{\cal L}_0$
and the perturbative one $\hat{\cal L}_1$ are scale invariant.
The bare zero modes scales similarly. And it is very essential that
the first perturbation step should dismiss the degeneracy:
on the first step of the iteration procedure
it should appear two zero modes
\begin{eqnarray}&&
{\cal Z}_1=\sum\biggl[\bigl(\alpha Z_{ij,kl}+\beta Z_{ij,ik}\bigr)
\bigl(1-\Delta_1\ln[r/L]\bigr)\biggr],\label{Z1}\\&&
{\cal Z}_2=\sum\biggl[\bigl((1-\alpha) Z_{ij,kl}-\beta Z_{ij,ik}\bigr)
\bigl(1-\Delta_2\ln[r/L]\bigr)\biggr],\nonumber\\&&
\label{Z2}
\end{eqnarray}
such that
\begin{equation}
{\cal Z}=\frac{(2-\gamma)^2}{\gamma^2d^4}({\cal Z}_1+{\cal Z}_2),
\label{Z}
\end{equation}
 where ${\cal Z}$ was found above in (\ref{zr}).

To extract an additional information that will fix all the numbers
$\alpha,\beta,\Delta_i$ we should calculate first-order logarithmic
correction $-\hat{\cal L}_0^{-1}\hat{\cal L}_1\sum{\cal Z}_{in,jn}$
to the second bare zero mode $\sum {\cal Z}_{in,jn}$.
First of all one finds
\begin{eqnarray}&&
-\hat{\cal L}_1\sum  {\cal Z}_{in,jn}=
\frac{(d-1)\gamma^2}{2(2-\gamma)}\sum \bigl(r_{ij}^2-r_{in}^2-r_{jn}^2\bigr)
r_{jn}^{\gamma-2}\nonumber\\&&+
\frac{\gamma^2}{4}\sum\frac{r_{in}^{\gamma-2}r_{jn}^{\gamma-2}}{r_{ij}^\gamma}
%% FOLLOWING LINE CANNOT BE BROKEN BEFORE 80 CHAR
\biggr[\frac{d+1-\gamma}{2-\gamma}r_{ij}^2\bigl(r_{in}^2+r_{jn}^2-r_{ij}^2\bigr)
\nonumber\\&&+
\frac{1}{2}\bigl(r_{ij}^2+r_{in}^2-r_{jn}^2\bigr)
\bigl(r_{ij}^2+r_{jn}^2-r_{in}^2\bigr)\biggr].
\label{L1G2}
\end{eqnarray}
Performing calculations analogous to what was done at the calculations of
the coefficients $C_0^{i}$
one finds that all the logarithmic terms appeared are proportional to the bare
zero mode $\sum {\cal Z}_{in,jn}$ only. It is a manifestation of the fact
that the functions on a triangle constitute
an invariant subspace of the full
operator $\hat{\cal L}$: $\hat{\cal L}$ acting on an arbitrary function
of three distances from a triangle $r_{in},r_{ij},r_{jn}$ produces
a function of the same three distances again.  Thus
the cancelation of a coefficient before the logarithm proportional
to another four-point zero mode will occurs on all the higher steps of the
iteration of $\sum {\cal Z}_{in,jn}$ too. Thus, the resulting zero mode
(say ${\cal Z}_1$ with $\alpha=0$)
is scale invariant with the following asymptotics of the anomalous exponent
at $d\to\infty$
\begin{equation}\Delta_1\to-\frac{(2-\gamma)(2+\gamma)}{2d},
\label{d1a}
\end{equation}
The asymptotics of all the rest coefficients
defining the second (mixed) scale-invariant
zero mode ${\cal Z}_2$ is
restored from (\ref{zr},\ref{Z1},\ref{Z2},\ref{Z},\ref{d1a})
\begin{equation}\Delta_2\to \frac{4(2-\gamma)}{d}
\quad \beta\to\frac{4}{10+\gamma},
\label{dinf}
\end{equation}
Note that the above $\hat{\cal L}_1^{-1}\hat{\cal L}_0$ iteration procedure
was not, strictly speaking, a direct $1/d$-expansion:
not only major terms $\sim d^2$ but also subleading ones $\sim d$
have been included into $\hat{\cal L}_0$.
That made it possible to have both unperturbed operator and perturbation
of the same (second) order.
To reinforce the results and obtain the explicit solution for
$\Gamma_1$ (not only it's logarithmic part as above), let us briefly
describe the direct $1/d$-procedure where
the main part of $\hat{\cal L}$ in the convective
interval is
the differential operator of the first order
\begin{equation}
\hat{\cal L}^\prime_0=
d^2\frac{D}{2-\gamma}\sum_{i>j} r_{ij}^{1-\gamma}\partial_{r_{ij}}.
\label{d1}
\end{equation}
yet we will directly account for the necessary boundary conditions while
integrating over characteristics below.
The zero modes of (\ref{d1}) which we are going to iterate are $d\to\infty$
limits
of $\sum {\cal Z}_{ij,kl}$ and $\sum {\cal Z}_{ij,ik}$:
${\cal Z}^\prime_0=\sum \bigl[2r_{ij}^\gamma r_{kl}^\gamma-
\bigl(r_{ij}^{2\gamma}+r_{kl}^{2\gamma}\bigr)\bigr]$ and
$\tilde{\cal Z}^\prime_0=\sum \bigl[2r_{ij}^\gamma r_{ik}^\gamma-
\bigl(r_{ij}^{2\gamma}+r_{ik}^{2\gamma}\bigr)\bigr]$.
To find the first-order corrections to the bare zero modes
one has to solve the differential equation
\begin{eqnarray}&&
\hat{\cal L}^\prime_0{\cal Z}^\prime_1=-\hat{\cal L}^\prime_1{\cal Z}^\prime_0,
\quad \hat{\cal L}^\prime_0\tilde{\cal Z}^\prime_1=
-\hat{\cal L}^\prime_1\tilde{\cal Z}^\prime_0,
\label{d3}\\&&
\hat{\cal L}^\prime_1=
d\frac{D}{2-\gamma}\biggl\{
 \sum_{i>j} \bigl(r_{ij}^{2-\gamma}\partial^2_{r_{ij}}-
\gamma r_{ij}^{1-\gamma}\partial_{r_{ij}}\bigr)-\nonumber\\&&
\frac{1}{2}\sum(r_{in}^2-r_{ij}^2-r_{jn}^2)
\frac{r_{ij}^{1-\gamma}}{r_{jn}}
\frac{\partial^2}{\partial r_{ij}\partial r_{jn}}-
\nonumber \\ &&
\frac{1}{4}\sum
\frac{r_{ij}^{2-\gamma}}{r_{im}r_{jn}}
\bigl(r_{in}^2+r_{jm}^2-r_{ij}^2-r_{mn}^2\bigr)
\frac{\partial^2}{\partial r_{im}\partial r_{jn}}\biggr\},
\label{d4}
\end{eqnarray}
where $\hat{\cal L}^\prime_1$ is the part of the
full operator (in the convective interval) proportional to $d$,
note that it stems both from $\hat{\cal L}_0$ and $\hat{\cal L}_1$.
The equations (\ref{d3}) are integrated by characteristics, for example
\begin{equation}
{\cal Z}^\prime_1=-\int dt \hat{\cal L}^\prime_1{\cal Z}^\prime_0
\biggl[r_{ij}^\gamma\to \frac{d^2\gamma D}{2-\gamma}t+B_{ij}\biggr],
\label{d5}
\end{equation}
where $B_{ij}$ should be considered as constants, and ${\cal Z}^\prime_1$
itself is defined
upto zero modes of the bare operator $\hat{\cal L}^\prime_0$.
The result of integration (\ref{d5}) at an arbitrary
$\gamma$ is bulky, we present here the explicit expression in the simplest
possible case of $\gamma=1$
\end{multicols}
\begin{eqnarray}&&{\cal Z}^\prime_1(\gamma=1)=\frac{4}{d}\sum_{i>j}r_{ij}^2-
\frac{1}{2d}\sum\biggl(\frac{(r_{im}-r_{ij})
[(r_{im}-r_{in})^2+(r_{im}-r_{jm})^2-(r_{im}-r_{ij})^2-
(r_{im}-r_{mn})^2]}{r_{im}-r_{jn}}\ln[L/r_{im}]+\nonumber\\&&
\frac{(r_{jn}-r_{ij})
[-(r_{jn}-r_{in})^2-(r_{jn}-r_{jm})^2+(r_{jn}-r_{ij})^2+
(r_{jn}-r_{mn})^2]}{r_{im}-r_{jn}}\ln[L/r_{jn}]\biggr).
\nonumber\\&&
\tilde{\cal Z}^\prime_1(\gamma=1)=-\frac{8}{d}\sum_{i>j}r_{ij}^2-
\frac{1}{d}\sum \frac{(r_{ij}-r_{ik})(r_{ij}-r_{kj})}{r_{ik}-r_{kj}}\biggl(
%% FOLLOWING LINE CANNOT BE BROKEN BEFORE 80 CHAR
(r_{ij}+r_{kj}-2r_{ik})\ln[L/r_{ik}]-(r_{ij}+r_{ik}-2r_{kj})\ln[L/r_{kj}]\biggr).
\nonumber
\end{eqnarray}
\begin{multicols}{2}
One can directly check that those are the solutions of (\ref{b7}) at the first
order
in $1/d$.
Being interested only in finding the overall scaling we can just put
all the separations under the logarithm to be the same. Generally,
one can extract
the logarithmic parts of ${\cal Z}^\prime_1,\tilde{\cal Z}^\prime_1$ directly
from
the r.h.s. of (\ref{d5}) putting formally the upper limit of the integral
to be very large and looking for the logarithmically divergent terms
\begin{eqnarray}&&
{\cal Z}^\prime_1\!=\!\frac{2(2\!-\!\gamma)}{d}
\bigl(2{\cal Z}^\prime_0-\tilde{\cal Z}^\prime_0\!\bigr)\ln\biggl({L\over
r}\biggr)
\!+\!
\mbox{non-log. terms},
\label{d6}\\&&\tilde{\cal Z}^\prime_1=
\frac{\gamma^2-4}{d}\tilde{\cal Z}^\prime_0\ln\biggl({L\over r}\biggr)\!+
\!\mbox{non-log. terms}.\label{d8}\end{eqnarray}
This is the direct analog of (\ref{zr}).
The formulas (\ref{d6},\ref{d8})
immediately give the leading $d\to\infty$ asymptotics (\ref{d1a},\ref{dinf})
obtained above from the general iteration procedure.

The structure of the resolvent (\ref{r3b}) at small $r$ determined by
(\ref{r4-5}) shows that for $\Gamma^{(1)}$ there exists the limit $\kappa\to
0$.
This property being reproduced at any step of the iteration procedure
enables one to construct a solution $\Gamma^{(co)}$ which is the limit of
$\Gamma$ at $\kappa\to 0$. The function $\Gamma^{(co)}$ is close to
$\Gamma$ in the convective interval and remains finite at any
$r_{ij}\to 0$. The behavior of $\Gamma^{(co)}$ is determined by
the estimation (\ref{h8}) where $r$ is the maximal value among $r_{ij}$.

We thus come to the conclusion that for
separations $r$ from the convective interval $r_d\ll r\ll L$
\begin{equation}
\Gamma(r)-\Gamma(0) \sim
P_2^2 r^{2\gamma}(L/r)^{\Delta}/D^2,\quad\Gamma(0) \sim P_2^2L^{2\gamma}/D^2
\label{h8} \end{equation}
up to dimensionless constants depending on $d$ and $\gamma$
and the anomalous dimension $\Delta$ is given by the largest exponent
($\Delta_2$)
which is asymptotically
\begin{equation}\Delta=\Delta_2=4\frac{2-\gamma}{d}+O(1/d^2)\ .
\label{deltaa}\end{equation}

Finally, we treat the case where all separations are in the
diffusion interval. The contribution to $\Gamma^{(1)}$ associated
with $r^\prime\gg r_d$ gives the constant $\sim L^{2\gamma}(L/r_d)^\Delta
P_2^2/D^2$.
The contribution associated with scales $r^\prime \lesssim r_d$
has to be analyzed carefully since
for all separations from the diffusion interval the typical integration time
$t$
in (\ref{h5}) is also characteristic of the diffusion region. We use
(\ref{r3a})
for ${\cal R}^{(p)}$ and (\ref{a12c}) for $f$. The estimate for
the $r$-independent contribution gives $\sim r_d^{2\gamma}P_2^2/D^2$.
To analyze a $r$-dependent contribution associated with $r^\prime \lesssim r_d$
it is convenient to return to the initial form (\ref{b7}) of the equation
for $\Gamma$. At the analysis we will not divide $\hat{\cal L}$ into
$\hat{\cal L}_0$ and $\hat{\cal L}_1$, the consequent conclusions are
consequently non-perturbative. In the diffusion region, $\hat{\cal L}$
is reduced to the sum of Laplacians. The r.h.s. of (\ref{b7b}) is proportional
to $r^{4-\gamma}$, that generates the forced part of the solution proportional
to $r^{6-\gamma}$. To find the principal $r$-dependent term one has to compare
the forced solution with zero modes of the sum of Laplacians.
All the modes are well known: they are constructed from even powers of
$r_{ij}$. There are no modes with second powers of $r_{ij}$
in $\Gamma$ since they are not invariant under permutations of points
${\bf r}_i$. The zero mode of $\hat{\cal L}$ of the fourth power
possessing the permutation symmetry:
\begin{equation}
\sum_{\{ij\}\neq\{nm\}}\biggl(
r_{ij}^2 r_{nm}^2-\frac{d}{2(d+2)}(r_{ij}^{4}+
r_{nm}^4)\biggr) \,,
\label{difz} \end{equation}
turns out to produce larger contribution than the forced term.
Thus, we conclude that the principal term in $\Gamma$ is of the forth
power in $r$. The parametric dependence of the coefficient at the power can
be easily established from a matching at $r\sim r_d$:
\begin{eqnarray} &&
\underline{{\rm at} \quad r\ll r_d}
\nonumber \\ &&
\Gamma-(C_1L^{2\gamma}+C_1^\prime r_d^{2\gamma})P_2^2/D^2
\sim r^4 r_d^{2\gamma-4}(L/r_d)^\Delta P_2^2/D^2.  \nonumber\\&&
\label{h10} \end{eqnarray}

The consideration of this Section shows that the degree
of non-Gaussianity increases as the distances decrease in the
convective interval and eventually comes to a constant when
the distances are in the diffusion interval. For example, the flatness factor
is proportional to $(L/r)^\Delta$ in the convective interval
$L\gg r\gg r_d$ and it is an $r$-independent constant of the order of
$(L/r_d)^\Delta$ in the diffusive interval $r_d\gg r$.

\section{Four-point objects with strongly different separations}
\label{sec:twopa}

In Section \ref{sec:four} we established the overall scaling behavior of the
four-point irreducible correlation function $\Gamma_{1234}$. In particular,
one may check that at any step of
the iteration procedure (\ref{h3})  the contributions to $\Gamma_{1234}$
associated with possible small values of the separation ratios are
negligible in comparison with the leading term. To show that the
summation over $n$ could not convert subleading terms into leading ones, here
we
check the self-consistency of our approach nonpertubatively in $1/d$
using the full operator $\hat{\cal L}$.
In the next subsection, we show
that if any separation $\rho$ is much less than other ones
then the $\rho$-dependent contribution to $\Gamma_{1234}$ is much less
than the leading term. This property enables one
to find such two-point objects as $\langle (\theta_1-\theta_2)^4 \rangle$
which can be extracted from $\Gamma_{1234}$ by fusing some points.
Besides, to establish the properties of the correlation functions
$\langle\langle\epsilon_1\theta_2^2\rangle\rangle$ and
$\langle\langle\epsilon_1\epsilon_2\rangle\rangle$ one should know
not only the limit of $\Gamma_{1234}$ for coinciding points but also
the dependence of $\Gamma_{1234}$ on distances between nearby points
in groups strongly separated from each other. It is the aim
of this section to obtain the dependence. Here, we shall use
perturbation theory with respect to the ratio between small and large
distances. Calculating the corresponding contributions to $\Gamma_{1234}$
one can obtain the two-point correlation functions of products of
different spatial derivatives. We shall demonstrate
how an absence of an $r_d$-related anomalous scaling in
$\Gamma_{1234}$
prescribes it for $\langle\langle\epsilon_1\theta_2^2
\rangle\rangle$ and $\langle\langle\epsilon_1\epsilon_2\rangle\rangle$.

\subsection{Four-Point Correlator with a small Separation}

Suppose that among separations $r_{ij}$ there is the separation $\rho$,
say $r_{34}$,  which is much smaller than other separations. Then the
main term in the operator $\hat{\cal L}$ (\ref{b4}) is associated with
derivatives over the vector ${\bbox\rho}={\bf r}_3-{\bf r}_4$.
This term can be written as
\begin{equation}
\hat{\cal L}_{\bbox\rho}\equiv
{\cal K}^{\alpha\beta}({\bbox\rho})
\nabla_{\rho}^\alpha\nabla_{\rho}^\beta
+2\kappa\triangle_{\rho} \,
\label{b9} \end{equation}
where $\nabla_{\rho}^\alpha\equiv \partial/\partial\rho_{\alpha}$.
To solve the equation (\ref{b7a}) for
$\Gamma_{1234}$ in this case, we can formulate  an iteration procedure where
$\hat{\cal L}_{{\bbox\rho}}$ is treated as the main contribution
to $\hat{\cal L}$. Namely a solution of (\ref{b7a}) can be
represented as
\begin{mathletters} \label{jp1}
\begin{eqnarray} &&
\Gamma_{1234}=\Gamma_0({\bf r}_{13}, {\bf r}_{23})
+\sum_{n=1}^\infty \Gamma_{n}({\bf r}_{13}, {\bf r}_{23}, {\bbox\rho})\,,
\label{jp1a} \\ &&
-\hat{\cal L}_{{\bbox\rho}}\Gamma_{n}=
\Phi_n({\bf r}_{13},{\bf r}_{23},\bbox{\rho})
\label{jp2} \\ &&
\Phi_1=\hat{\cal L}\Gamma_0({\bf r}_1,{\bf r}_2)+
\Phi_{12;34}+\Phi_{13;24}+\Phi_{14;23},
\label{jp1b} \\ &&
\Phi_n=(\hat{\cal L}-\hat{\cal L}_{{\bbox\rho}})
\Gamma_{(n-1)},\quad n>1\,,
\label{jp1c} \end{eqnarray} \end{mathletters}

\noindent
where ${\bf r}_{13}$ and ${\bf r}_{23}$ are separations much larger than
${\bbox\rho}$ and $\Gamma_0$ is $\rho$-independent part of $\Gamma$.
The procedure introduced by (\ref{jp1}) can be considered as a series over the
small parameters $\rho/r_{13}$, $\rho/r_{23}$. We impose two boundary
conditions
on $\Gamma_{n}$: to be of order of $\Gamma_0$ at $\rho\sim r_{13},r_{23}$
and to remain finite at $\rho=0$.

The
r.h.s. of (\ref{jp2}) can be assumed to be known. A solution of (\ref{jp2})
satisfying the imposed boundary conditions has the following form
\begin{eqnarray} &&
\Gamma_{n}({\bf r}_1,{\bf r}_2,{\bbox\rho})=
% \nonumber \\ &&
\int_0^\infty dt \int d{\bf r}
{\cal R}(t;{\bbox\rho},{\bf r})
\Phi_n({\bf r}_1,{\bf r}_2,{\bf r}) \,,
\label{jp3} \end{eqnarray}
where ${\cal R}$ is the kernel of the resolvent
of the operator $\hat{\cal L}_{\bbox\rho}$:
\begin{equation}
(\partial_t-\hat{\cal L}_{{\bbox\rho}})
{\cal R}(t;\bbox{\rho},{\bf r})=
\delta(t)\delta({\bf r}-\bbox{\rho}) \,.
\label{L3} \end{equation}
Since the operator $\hat{\cal L}_{{\bbox\rho}}$ explicitly depends on
${\bbox\rho}$ the resolvent ${\cal R}$ is not a function of the difference
$\mid {\bf r}-\bbox{\rho}\mid$ only as it would be in a homogeneous case.
The symmetry admits to treat ${\cal R}$ as a function of $\rho$, $r$ and
$x={\bbox\rho}{\bf r}/\rho r$. It is convenient to expand the function
in the series over Jacobi polynomials \cite{Korn} which are the eigenfunctions
of the $x$-dependent part of $\hat{\cal L}_{{\bbox\rho}}$:
\begin{equation}
{\cal R}(\rho,r,x)=\sum_{n=0}^\infty
{\cal R}^{(2n)}(\rho,r)
P_{2n}^{(\nu,\nu)}(x) \,,
\label{ll4} \end{equation}
where $\nu=(d-3)/2$ and only the polynomials with
even numbers participate since ${\cal R}$ is an even function of $x$
(at $d=3$, Jacobi polynomials turn into Legendre ones).
One can obtain from (\ref{L3}) separate equations for ${\cal R}^{(2n)}$:
\begin{eqnarray} &&
\Biggl(\partial_t -\hat{\cal L}^{(p)}_\rho+
\biggl(\frac{d+1-\gamma}{2-\gamma}D\rho^{2-\gamma}+2\kappa\biggr)
\frac{2n(2n+d-2)}{\rho^2}\Biggr)
\nonumber \\ &&
\times {\cal R}^{(2n)}(\rho,r)=\delta(\rho-r)\rho^{-(d-1)}\times
\nonumber \\ &&
\frac{(d-2+4n)\Gamma(d-2+2n)}
{2^{d-1}\pi^{(d-1)/2}\Gamma((d-1)/2+2n)} \,,
\label{ll5} \end{eqnarray}
where $\hat{\cal L}^{(p)}_\rho$ was introduced in (\ref{a9b}).

Solving (\ref{ll5}) one gets the behavior of ${\cal R}^{(2n)}$
in the diffusive and convective intervals explicitly:
\begin{mathletters} \label{pn2}
\begin{eqnarray} &&
\underline{r,\rho,\sqrt{t\kappa}\ll r_d}\quad
{\cal R}^{(2n)}=\frac{(d-2+4n)\Gamma(d-2+2n)}
{2^{d+1}\pi^{(d-1)/2}\Gamma(2n-1+d/2)}
\nonumber \\ &&
\times\exp\biggl(-\frac{r^2+\rho^2}{8t\kappa}\biggr)
I_{d/2- 1+2n}\biggl(\frac{r\rho}{4t\kappa}\biggr)
\bigl(r\rho\bigr)^{1-d/2},
\label{pn2ab} \\ &&
\underline{r,\rho,(Dt)^{1/\gamma}\gg r_d}\quad
{\cal R}^{(2n)}=
\label{pn2cd} \\ &&
\frac{(2-\gamma)(d-2+4n)\Gamma(d-2+2n)}
{\gamma(d-1)2^{d-1}\pi^{(d-1)/2}\Gamma(2n-1+d/2)}
\nonumber \\ &&
\times \exp\biggl(-\frac{(r^\gamma+\rho^\gamma)(2-\gamma)}
{\gamma^2(d-1)tD}\biggr)
I_{\eta_n}\biggl(\frac{2(2-\gamma)(r\rho)^{\gamma/2}}
{\gamma^2(d-1)tD}\biggr) \,,
\nonumber \end{eqnarray} \end{mathletters}
where
\begin{equation}
\eta_n=\frac{1}{\gamma}\sqrt{(d-\gamma)^2+
\frac{8(d+1-\gamma)n(2n+d-2)}{d-1}}\,,
\label{eta} \end{equation}
and $I_\eta$ designate modified Bessel functions.
Note that ${\cal R}^{(0)}={\cal R}^{(p)}{\Gamma(d/2)}r^{1-d}/(2\pi^{d/2})$
where ${\cal R}^{(p)}$ was introduced by (\ref{r1},\ref{r3}).
It is possible to formulate for ${\cal R}^{(2n)}$ interpolation formulas
of the (\ref{r4-5}) type.

To establish the behavior of $\Gamma_n$ determined by (\ref{jp1})
we should analyze the integral (\ref{jp3}) using (\ref{pn2}). The
main part of the integral is determined by the region $r\sim r_{13},r_{23}$.
Thus the $\rho$\,-dependence is associated with the limit of (\ref{pn2})
at $\rho\ll r$. We see that the resolvents in this case
tend to zero with decreasing $\rho$. We can also assert based on
(\ref{eta}) that the higher is the number $n$ of the angular harmonic
the faster is the decay of the corresponding resolvent. Since the procedure
(\ref{jp1}) produces a convergent series over a small parameter then the above
statements can be extended to the sum of the series. Thus we proved
that the $\rho$-dependent part of $\Gamma$ is parametrically smaller
than its $\rho$-independent part at $\rho\ll r_1,r_2$ what was the
purpose of this subsection.

\subsection{Four-Point Correlator with two separated pairs}
\label{subsec:IP}

At the rest of this section, we consider $\Gamma_{1234}$ in
the case of the special geometry with two separations between points
being much smaller than all other ones, namely $R\gg\rho_{1,2}$, where
\begin{eqnarray} &&
{\bf r}_1-{\bf r}_2=\bbox{\rho}_1,\qquad {\bf r}_3-{\bf r}_4=\bbox{\rho}_2,
\nonumber \\ &&
{\bf R}=({\bf r}_3+{\bf r}_4-{\bf r}_1-{\bf r}_2)/{2}\,.
\label{b8} \end{eqnarray}
Here, the mutual orientations of the vectors and the ratio $\rho_1/\rho_2$
are supposed to be arbitrary. We
denote by $\Gamma_0(R)$
the main contribution to $\Gamma_{1234}$  which is equal
to $\Gamma_{1234}$ at $\rho_1=\rho_2=0$.
The estimate for $\Gamma_0$ can be extracted from the results that have been
obtained in
Section \ref{sec:four} and justified in the preceding subsection.
Indeed, the statements (\ref{h8},\ref{h10}) are valid for $r$ being the
largest length $R$ of the tetrahedron. One thus gets for $\Gamma_0$:
\begin{mathletters} \label{bb}
\begin{eqnarray} &&
\underline{r_d \ll R \ll L} \quad
\Gamma_0\approx (C_1 L^{2\gamma}-C_2 R^{2\gamma}(L/R)^\Delta)P_2^2/D^2 \,,
\nonumber\\&&
\label{bb2} \\ &&
\underline{R \ll r_d} \quad
\Gamma_0\approx (C_1 L^{2\gamma}
-C_3 R^{4} (L/R)^\Delta r_d^{2\gamma -4})P_2^2/D^2 \,,\nonumber\\&&
\label{bb3} \end{eqnarray} \end{mathletters}

\noindent
where $C_1$, $C_2$, $C_3$ are dimensionless constants depending on $d$ and
$\gamma$.

Generally, we do not assume here and below in this Section
that $\Delta$ is small (or $d$ is large)
yet the condition $\Delta<2\gamma$ is assumed.
At the limit $d\rightarrow\infty$, the numerical coefficients $C_1,C_2,C_3$
could
be found from the results of Sect. II.
In particular, we can get the following asymptotic expressions:
\begin{eqnarray}&&
\langle(\theta_1-\theta_2)^4\rangle\approx
\frac{24(2-\gamma)^2}{\gamma^2d^4}r^{2\gamma}\biggl[
(1-2\beta)\biggl(\frac{L}{r_{12}}\biggr)^{\Delta_2}\nonumber\\&&+
2\beta\biggl(\frac{L}{r_{12}}\biggr)^{\Delta_1}\biggr]+\cdots,
\label{tmta}\\&&
\mbox{at}\quad r_{12}\ll r_{13}\approx r_{23}\quad
\langle(\theta_1-\theta_2)^2\theta_3^2\rangle-
\langle(\theta_1-\theta_2)^2\rangle
\langle\theta_3^2\rangle\nonumber\\&&\approx
\frac{4(2-\gamma)^2\beta r_{12}^\gamma r_{13}^\gamma}{\gamma^2d^4}
\biggl[-\biggl(\frac{L}{r_{13}}\biggr)^{\Delta_2}\!\!+
\biggl(\frac{L}{r_{13}}\biggr)^{\Delta_1}\biggr]+
\cdots,\nonumber\\&&\label{tmtb}\\&&
\mbox{at}\quad r_{12},r_{34}\ll r_{13}\approx r_{23}\approx r_{24}\approx
r_{14}\nonumber\\&&\approx
\langle(\theta_1-\theta_2)^2(\theta_3-\theta_4)^2\rangle-
\langle(\theta_1-\theta_2)^2\rangle
\langle(\theta_3-\theta_4)^2\rangle\approx\nonumber\\&&
\frac{4(2-\gamma)^2}{\gamma^2(d-1)^2d^2} r_{12}^\gamma r_{34}^\gamma
\biggl(\frac{L}{r_{13}}\biggr)^{\Delta_2}+\cdots,\label{tmtc}
\end{eqnarray}
where dots designate the terms with a normal scaling
($r^{2\gamma}$). Note that
if one expands (\ref{tmta}) in small $\Delta$-s then the remarkable
cancelation of the terms linear in logs happens: $(1-2\beta)\Delta_2+
2\beta\Delta_1=0$.

Now, we aim at finding leading corrections to $\Gamma_0(R)$ at small
values of $\rho_1/R$ and $\rho_2/R$. For this we should solve (\ref{b7a})
at the conditions $\rho_1,\rho_2\ll R$. In this limit the principal part
of $\hat{\cal L}$ is $\hat{\cal L}_{{\bbox\rho}_1}+\hat{\cal
L}_{{\bbox\rho}_2}$
where $\hat{\cal L}_{\bbox\rho}$ is determined by (\ref{b9}).
It is the reason which caused us to formulate an iteration procedure where
$\hat{\cal L}_{{\bbox\rho}_1}+\hat{\cal L}_{{\bbox\rho}_2}$ is treated as a
main contribution to $\hat{\cal L}$. Namely a solution of the equation
(\ref{b7a}) can be represented as
\begin{mathletters} \label{ip1}
\begin{eqnarray} &&
\Gamma_{1234}=\Gamma_0(R) +\sum_{n=1}^\infty
\Gamma_{n}({\bf R};{\bbox\rho}_1,{\bbox\rho}_2) \,,
\label{ip1a} \\  &&
-(\hat{\cal L}_{{\bbox\rho}_1}+\hat{\cal L}_{{\bbox\rho}_2})\Gamma_{n}=
\Phi_n({\bf R};\bbox{\rho}_1,\bbox{\rho}_2)
\label{ip2}\\&&
\Phi_1=\hat{\cal L}\Gamma_0+
\Phi_{12;34}+\Phi_{13;24}+\Phi_{14;23},
\label{ip1b} \\ &&
\Phi_n=(\hat{\cal L}-\hat{\cal L}_{{\bbox\rho}_1}-
\hat{\cal L}_{{\bbox\rho}_2})\Gamma_{(n-1)},\quad n>1.
\label{ip1c}
\end{eqnarray} \end{mathletters}

\noindent
The procedure introduced by (\ref{ip1}) can be considered as a series over the
small parameters $\rho_1/R$, $\rho_2/R$. We impose two conditions
on $\Gamma_{n}$: to be of order of $\Gamma_0$ at $\rho_1,\rho_2\sim R$
and to turn into zero at $\rho_{1,2}=0$.

Before consistent derivation, let us notice that the $r_d$-dependence
could be readily found from (\ref{ip2}) and the additional anomalous scaling of
the subleading ($\rho$-dependent) terms naturally appears already at the first
step as a consequence of the given scaling of $\Gamma_0$. Let us consider
$\rho\ll r_d$, then the solution of (\ref{ip2}) for $\Gamma_1$ will be
$\rho_1^2+\rho_2^2$ multiplied by the $R$-dependent terms $\Phi$ and
$\hat{\cal L}\Gamma_0$ which both are proportional to $R^\gamma(L/R)^\Delta$
-- see (\ref{g11}) below. Assuming that
no special cancelations happen, one may conclude, in particular, that
$\langle\langle\theta^2(0)\epsilon(R)\rangle\rangle\propto r_d^{\gamma-2}
R^\gamma (L/R)^\Delta$ instead
of $R^{2\gamma-2} (L/R)^\Delta$ that would be given by a naive counting of
powers.

Let us start now a regular examination of (\ref{ip2}).
A solution of that equation has the following form
\begin{eqnarray} &&
\Gamma_{n}({\bf R};\bbox{\rho}_1,\bbox{\rho}_2)=\int_0^\infty dt
\label{ip3} \\ &&
\times\int\int d{\bf r}^\prime d{\bf r}^{\prime\prime}
{\cal R}(t;\bbox{\rho}_1,{\bf r}^\prime)
{\cal R}(t;\bbox{\rho}_2,{\bf r}^{\prime\prime})
\Phi_n({\bf R};{\bf r}^\prime,{\bf r}^{\prime\prime}) \,,
\nonumber \end{eqnarray}
where ${\cal R}$ is the kernel of the resolvent of the operator
$\hat{\cal L}_{\bbox\rho}$ introduced by (\ref{L3},\ref{ll4},\ref{pn2}).
The main contribution to the integral (\ref{ip3}) is determined by
the region $r^\prime\sim R$, $r^{\prime\prime}\sim R$.
The higher is the number of the angular harmonic $n$ in
(\ref{ll4},\ref{pn2}) the smaller is the respective contribution
into $\Gamma$ due to additional powers of the small parameter $\rho/R$
in ${\cal R}^{(2n)}$. Therefore besides the first harmonic,
we consider further only the second one to show how a non-trivial
anomalous exponent appears due to an angular dependence.

\subsection{Zero modes}
\label{subsec:ZF}

The asymptotic formulas (\ref{pn2}) for the resolvent enable one to
extract a scaling behavior of the general expression (\ref{ip3})
for a solution of (\ref{ip2}). We consider the case where the largest
scale $R$ of the tetrahedron is much smaller than the pumping scale $L$.
That means that
the only external scales for the integrations in (\ref{ip3}) can be $R$
or $r_d$ but not $L$. To be interested in the contribution to
$\Gamma_{n}$ stemmed from large enough scales
$r^\prime,r^{\prime\prime}\sim R$ and time $t\sim\tau_R$
(where one defines $\tau_r$ as $r^2/\kappa$ at $r\ll r_d$ but as
$r^\gamma/D$ at $r\gg r_d$), one can expand the resolvents in
(\ref{ip3}) in a series over the small parameters
$\tau_r/t$ and analyze the convergence of the temporal integral for
different terms of the expansion. Of course, the integrals depend on the
concrete form of $\Phi$. However, in any case only a finite number of the
first terms of the expansion are determined by the time
$t\sim\tau_R$. The respective contributions into $\Gamma$ are the
zero modes of the operator $\hat{\cal L}_{{\bbox\rho}_1}+
\hat{\cal L}_{{\bbox\rho}_2}$ that grow with $\rho$ and are
symmetric under the permutation $\rho_1\leftrightarrow\rho_2$.
All the rest terms (stemmed from $r^\prime,r^{\prime\prime}\ll R$)
are summed up into the term estimated as $\tau_\rho\Phi$, we will
name this part of the solution the forced one $\Gamma^f$.

The $r$-dependence of the forced part of the term $\Gamma_{n}$
depends on a concrete form of the function $\Phi_n$ in (\ref{ip2}).
Contrary, the form of the zero modes is universal: they
depend on the source term $\Phi_n$ only via coefficients. We thus start
from finding out the zero modes that give the main contributions at small
scales. To find the zero modes that contribute to the solution, we will
follow an indirect (but, probably, the simplest) way. Since the zero
modes are formed at the large times $t\sim\tau_R$ in the integral (\ref{ip3})
(which means that they appear due to matching at $\rho\sim R$). then
we can extract their scaling behavior expanding the expressions (\ref{pn2})
for the resolvents in the series over $\tau_\rho/t$ (\ref{pn2}) and keeping
the first terms of the expansion. Thus we can extract the exponents of the
first zero modes and their angular dependence but not coefficients at
different terms of the same order. To find the coefficients we can
construct a combination of a given order with arbitrary coefficients
and then find relations between the coefficients
demanding that the combination is a zero mode of
$\hat{\cal L}_{{\bbox\rho}_1}+\hat{\cal L}_{{\bbox\rho}_2}$, what
can be established by direct applying the operator.

Let us realize the scheme using the expressions (\ref{pn2}). We see
that the $\rho$-expansion of the expressions produces powers of
$\rho^\gamma$ and $\rho^\delta$ where as follows from (\ref{eta})
\begin{equation}
\delta= \frac{1}{2}\biggl(\gamma-d+
\sqrt{(d-\gamma)^2+\frac{8(d+1-\gamma)d}{d-1}}\biggr)\,.
\label{delta} \end{equation}
The angular structure of the zero modes also can be established if
to take into account the explicit form of two first Jacobi polynomials:
\begin{eqnarray}
P_0^{(\nu,\nu)}(x)=1\,, \quad
P_2^{(\nu,\nu)}(x)=\frac{d+1}{8}(dx^2-1) \,.
\nonumber \end{eqnarray}
Then one can directly check that the zero modes of the first and of
the second orders are as follows:
\end{multicols}
\begin{mathletters} \label{ZM}
\begin{eqnarray}&&
{\cal Z}_2^{(0,0)}\sim \frac{P_2^2}{D^2}
\biggl(\frac{L}{R}\biggr)^\Delta
\Biggl\{
\begin{array}{c} \rho_1^\gamma\rho_2^\gamma-
\frac{d}{2(d+\gamma)}(\rho_1^{2\gamma}+\rho_2^{2\gamma}),
\quad \rho\gg r_d,\\
%% FOLLOWING LINE CANNOT BE BROKEN BEFORE 80 CHAR
\bigl(\rho_1^2\rho_2^2-\frac{d}{2(d+2)}(\rho_1^4+\rho_2^4)\bigr)r_d^{2\gamma-4},
\quad \rho\ll r_d;\,
\end{array}\label{ZMa} \\ &&
{\cal Z}_1^{(0,2)}\sim
\frac{P_2^2}{D^2} R^{2\gamma-\delta}\biggl(\frac{L}{R}\biggr)^\Delta
\Biggl\{ \begin{array}{c}
\bigl(\{\rho_1^{\delta-2}(d({\bf R}\bbox{\rho}_1)^2- R^2\rho_1^2)\}+
\{\bbox{\rho}_1
\leftrightarrow\bbox{\rho}_2\}\bigr)R^{-2},
\quad \rho\gg r_d,\\
\bigl(\{d({\bf R}\bbox{\rho}_1)^2- R^2\rho_1^2\}+ \{\bbox{\rho}_1
\leftrightarrow\bbox{\rho}_2\}\bigr)r_d^{\delta-2}R^{-2},
\quad \rho\ll r_d;\, \end{array}
\label{ZMb} \\ &&
{\cal Z}_2^{(0,2)}\sim
\frac{P_2^2}{D^2} R^{\gamma-\delta}\biggl(\frac{L}{R}\biggr)^\Delta
\Biggl\{ \begin{array}{c}
\{\bigl(\rho_2^\gamma-b_1\rho_1^\gamma)\rho_1^{\delta-2}
\bigl(d({\bf R}\bbox{\rho}_1)^2- R^2\rho_1^2\bigr)\}R^{-2}+
\{\bbox{\rho}_1\leftrightarrow\bbox{\rho}_2\},
\ \rho\gg r_d,\\
\bigl[\{\bigl(\rho_2^2-b_2\rho_1^2)
\bigl(d({\bf R}\bbox{\rho}_1)^2- R^2\rho_1^2\bigr)\}R^{-2}+
\{\bbox{\rho}_1\leftrightarrow\bbox{\rho}_2\}\bigr]r_d^{\gamma+\delta-4},
\ \rho\ll r_d;\,
\end{array}\label{ZMb-c}\\&&
{\cal Z}_2^{(2,2)}\!\sim\!
\frac{P_2^2}{D^2} R^{2(\gamma-\delta)}\biggl(\frac{L}{R}\biggr)^\Delta
\Biggl\{ \begin{array}{c}
\bigl(\!\{\rho_1^{\delta-2}(d({\bf R}\bbox{\rho}_1)^2- R^2\rho_1^2)\}
\{\bbox{\rho}_1\!\leftrightarrow\!\bbox{\rho}_2\}\bigr)R^{-4}\!+\!c_1
(\rho_1\rho_2)^{\delta-2}
\bigl(d(\bbox{\rho}_1\bbox{\rho}_2)^2-\rho_1^2\rho_2^2\bigr),
\ \rho\gg r_d,\\
\bigl[\bigl(\{d({\bf R}\bbox{\rho}_1)^2- R^2\rho_1^2\}
\{\bbox{\rho}_1\leftrightarrow\bbox{\rho}_2\}\bigr)R^{-4}\!+\!c_2
%% FOLLOWING LINE CANNOT BE BROKEN BEFORE 80 CHAR
\bigl(d(\bbox{\rho}_1\bbox{\rho}_2)^2-\rho_1^2\rho_2^2\bigr)\bigl]r_d^{2\delta-4},
\ \rho\ll r_d;\,
\end{array}\nonumber\\&&\label{ZMc}
\end{eqnarray}
\end{mathletters}
\begin{multicols}{2}
\noindent
where $c_{1,2}$ are dimensionless constants and
\begin{equation}
b_1=\frac{(\delta+\gamma)(d+2\gamma-2)-2d}{d\gamma},\quad
b_2=\frac{d+4}{d} \,.
\nonumber \end{equation}
The upper indices of
${\cal Z}$, introduced in (\ref{ZM}), characterize the type of a zero mode
with respect to the angular structure. The lower index denotes the original
order
of the resolvent's expansion over $\tau_\rho/t$ producing the concrete term.
We keep the terms of the second order (\ref{ZMa},\ref{ZMb-c},\ref{ZMc})
besides the first order term (\ref{ZMb}) since they possess qualitatively
different $\rho$-dependence: the term (\ref{ZMb}) is an additive one while
the terms (\ref{ZMa},\ref{ZMb-c},\ref{ZMc}) contain cross-contributions.

Note that higher-order angular harmonics will be characterized by the
exponents $\delta_n$ which can be extracted from the asymptotic
behavior of the resolvents (\ref{pn2}) characterized by (\ref{eta}).
Let us write the explicit expressions for the exponents:
\begin{equation}
\delta_n\!=\! \frac{1}{2}
\left(\gamma-d\!+\!\sqrt{
(d-\gamma)^2\!+\!\frac{8n(d+1-\gamma)(2n+d-2)}{d-1}}\right)
\nonumber \end{equation}
which at $n=1$ gives (\ref{delta}). The exponents $\delta_n$ will
figure in the expressions of the (\ref{ZM}) type for higher harmonics.
It is worthwhile to emphasize here that the exponents $\delta_n$ are
nontrivial:
to obtain them by
$\hat{\cal L}_0^{-1}\hat{\cal L}_1$-expansion that we used in Section
\ref{sec:four}:
one would get terms like $\rho^\gamma\ln(\rho/R)$ that should be summed up into
$\rho^\delta$.

We introduced in (\ref{ZM}) the dimensional factors (in front of
the braces) with which the zero modes appear at $\Gamma$.
The factors in front of the braces are taken from the analysis of the
 integral (\ref{ip3}); since the contributions stem from the region
$\rho\sim R$ (where the resolvent is not precisely known)
then the dimensionless coefficients (including $c_1$ and $c_2$)
could not be found within our approach with one exception that will
be described below in Subsection \ref{subsec:Com}.
The dependence of the zero modes on
$r_d$ can be established by matching the respective term from
inertial and diffusive interval at $r_d$ and using the $r_d$-independence
of $\Gamma$ in the convective interval.

\subsection{Forced Terms}
\label{subsec:FT}

Let us begin the discussion of the forced terms with $\Gamma_{1}^f$ that
appears at the first step of the iteration procedure (\ref{ip1b}).
The r.h.s. of (\ref{ip1b}) contains the terms of different order in $\rho/ R$,
different angular functions and different types of the dependence on $\rho_1$
and $\rho_2$ (additive and multiplicative). We shall analyze the respective
contributions order by order in $\rho/ R$ separately for the different angular
functions. The terms with the cross-dependence (like $\rho_1^2\rho_2^2$)
are of special interest since they contribute into
$\langle\epsilon\epsilon\rangle$
correlation function.

The leading in $\rho/R$ contribution in r.h.s. of (\ref{ip1b}) is
$\rho$-independent
(and consequently angular independent). The contribution originates

\noindent
a) from $\hat{\cal L}\Gamma_0$:
\begin{mathletters} \label{g0}
\begin{equation}
\hat{\cal L}\Gamma_0
\rightarrow (\hat{\cal L}_R-\kappa\triangle_R)\Gamma_0(R);\label{g0a}
\end{equation}

\noindent
b) from $\Phi_{13;24}+\Phi_{14;23}$:
\begin{equation}
\label{g0b}
\Phi_{13;24}+\Phi_{14;23}\rightarrow \Phi_R\equiv 2
{\cal K}^{\alpha\beta}_R\frac{R^\alpha R^\beta}{R^2}
\biggl(\frac{df(R)}{dR}\biggr)^2.
\end{equation}
\end{mathletters}

\noindent
The expression for the respective forced solution is
\begin{eqnarray}&&
\Gamma_{1,add}^f\approx
\left[-\frac{(2-\gamma)}{(d-1)dD}
\Phi_R+(\hat{\cal L}_R-\kappa \triangle_{\bf R})\Gamma_0(R)\right]\nonumber\\&&
\times\biggl(\int^{\rho_1}_0
\frac{rdr}{2r_d^{2-\gamma}+r^{2-\gamma}}+
\int^{\rho_2}_0
\frac{rdr}{2r_d^{2-\gamma}+r^{2-\gamma}}\biggr) \,.
\label{g11} \end{eqnarray}
The subsequent additive terms are produced by taking into account
the next $\rho^2$-terms at the r.h.s.
of (\ref{ip1b}). They produce the additive forced contributions
to $\Gamma_{1}$ with the scaling $\propto\rho^{2+\gamma}$ in the
convective interval.

The leading cross-term originates from $\Phi_{12;34}$,
the main term of its expansion is
\begin{equation}
\Phi_{12;34}\approx
\nabla^\nu_{\bf _R}\nabla^\mu_{\bf _R}{\cal K}_{\bf _R}^{\alpha\beta}
\frac{\rho_1^\nu\rho_1^\alpha\rho_2^\beta\rho_2^\mu}{\rho_1\rho_2}
\frac{df(\rho_1)}{d\rho_1}\frac{df(\rho_2)}{d\rho_2}\,.
\label{per3} \end{equation}
Apart from (\ref{per3}), we have to take into account the cross-terms
originating from other terms in r.h.s. of (\ref{ip1b}). The terms have
a character of a regular expansion in $\rho/R$ and are consequently
proportional to $\rho_1^2\rho_2^2$. Such terms are much less than
(\ref{per3}) in the convective interval and are of the same order in the
diffusive interval. Since we are interested not in the factors but only in
the scaling behavior we can restrict ourselves with the analysis of
(\ref{per3}). The term is proportional to $\rho^{2\gamma}$ in the convective
interval and to $\rho^{4}$ in the diffusive one, that gives for the principal
behavior of the corresponding forced term $\rho^{3\gamma}$ and
$\rho^{6}$ respectively.

The source (\ref{per3}) produces not only cross-terms in $\Gamma$
of scalar and tensor structures but also additive and generally mixed ones
possessing all the variety of angular and
$\bbox{\rho}_1\leftrightarrow\bbox{\rho}_2$ symmetries used in the
classification
of the zero modes (\ref{ZM}). All those terms scale with $R$ as $R^{-\gamma}$
both in the convective and diffusive intervals by $\rho$.
To match those forced terms from the convective and diffusive intervals
one has to introduce the additional zero modes in the diffusive interval by
$\rho$
\begin{mathletters}
\label{ZN}
\begin{eqnarray}&&
{\cal Z}^{(0,0)}_{2,d}\sim {\cal Z}^{(0,0)}_2\bigl(R/L\bigr)^\Delta
\bigl({r_d}/{R}\bigr)^\gamma,
\label{ZNa}\\&&
{\cal Z}^{(0,2)}_{1,d}\sim {\cal Z}^{(0,2)}_1\bigl(R/L\bigr)^\Delta
\bigl({r_d}/{R}\bigr)^{3\gamma-\delta},
\label{ZNb}\\&&
{\cal Z}^{(0,2)}_{2,d}\sim {\cal Z}^{(0,2)}_2\bigl(R/L\bigr)^\Delta
\bigl({r_d}/{R}\bigr)^{2\gamma-\delta},
\label{ZNb-c}\\&&
{\cal Z}^{(2,2)}_{2,d}\sim {\cal Z}^{(2,2)}_2\bigl(R/L\bigr)^\Delta
\bigl({r_d}/{R}\bigr)^{3\gamma-2\delta}.
\label{ZNc}
\end{eqnarray}
\end{mathletters}

Let us now consider the forced term $\Gamma_{2}^f$ arising at the second
step of the iteration procedure. Analyzing the structure of the r.h.s. of
(\ref{ip1c}) one concludes that it
decreases downscales  with $\rho$ not slower than ($\Gamma_{1}^f+$zero modes).
Consequently, the forced term $\Gamma_{2}^f$ decreases downscales with $\rho$
not slower than $\tau_\rho(\Gamma_{1}^f+$zero modes) that is faster than
$\Gamma_{1}^f$. Thus, one can drop $\Gamma_{2}^f$ in comparison with
$\Gamma_{1}^f$ and all the set of zero modes. The same arguments allows one to
drop
the higher forced terms $\Gamma_{n}^f$, $n>2$ too.

\subsection{Comparison between the zero modes and forced terms}
\label{subsec:Com}

To find the main contributions into $\Gamma$ one should compare the zero modes
and
forced terms of the same angular structure.
The leading term in the expansion in $\rho/R$ is the isotropic additive forced
term (\ref{g11}) inside of both the convective and diffusive intervals.

The zero modes are $L$-dependent but the non-additive forced
terms are not. Thus keeping the parameter $L/R$ to be large enough one forces
the zero modes to dominate. A possibility for the forced terms to prevail
appears at the smallest $\rho$-separations, when the dominance of
the zero modes could be compensated by different $\rho$-dependencies
of the zero modes and the forced terms.
Present Subsection is devoted to such a comparison.

We compare (\ref{ZMa}) with the forced terms generated by (\ref{per3}).
Both in the convective and diffusion
interval we get the dominance of the angular independent zero mode
(\ref{ZMa}) containing the cross-contribution.
To be sure that (\ref{ZMa}) does give the main
contribution into $\langle\langle\epsilon_1\epsilon_2\rangle\rangle$
(and thus provides for its anomalous scaling),
we have to check whether or not the respective dimensionless
coefficient in front of that mode can turn into zero. We consider that mode in
a special case $R\gg L\gg\rho$ when the coefficient can be found unambiguously
contrary to other cases where the matching at $\rho\sim R$ is necessary. If
$R\gg L\gg\rho$ then one have to consider the integral (\ref{ip3}) with
(\ref{per3}) and the respective term from (\ref{pn2})
of the expansion of ${\cal R}^{(0)}$ -- see (\ref{r3b},\ref{r4-5}).
One can directly check that the main contribution into the integral
stems from the scales $L$ and time $\tau_L$ so that no matching at $R$
is necessary. We do not write here the bulky expression for the integral, it
turns into zero only at a single value $\gamma/d\approx0.086$ found
numerically.
That numerical value depends on the shape of $\chi$. At the rest values of
$\gamma/d$ the integral is nonzero and provides the contribution of the
structure (\ref{ZMa}).

Next we consider the second angular harmonic presented in
(\ref{ZMb},\ref{ZNb}),
(\ref{ZMb-c},\ref{ZNb-c})
and the product of the second angular harmonics in (\ref{ZMc},\ref{ZNc}).
The scaling of the respective harmonics of the forced terms coincides
with that of the isotropic contribution: $\rho^{3\gamma}$ and
$\rho^{6}$ in the convective and diffusive intervals respectively.
The zero modes (\ref{ZMb},\ref{ZMb-c}) and (\ref{ZMc})  scale
as $\rho^{\delta}, \rho^{\delta+\gamma}$ and $\rho^{2\delta}$
in the convective interval,
and as $\rho^2,\rho^2$ and $\rho^4$ in the diffusive one respectively.
One concludes:

\noindent
a) in the convective interval, the zero mode ${\cal Z}_1^{(0,2)}$
(${\cal Z}_2^{(0,2)}$ or ${\cal Z}_2^{(2,2)}$) is dominant if $3\gamma>\delta$
(respectively $2\gamma>\delta$ or $\gamma>2\delta/3$),
otherwise at the smallest $\rho$ the forced terms prevail;

\noindent
b) in the diffusive interval, at $3\gamma>\delta$
($2\gamma>\delta$ or $\gamma>2\delta/3$) the zero mode (\ref{ZMb})
[respectively (\ref{ZMb-c}) or (\ref{ZMc})]
become dominant at small enough $\rho$; otherwise the zero mode (\ref{ZNb})
[respectively (\ref{ZNb-c}) or (\ref{ZNc})],
matching with the forced solution originating from (\ref{per3}), prevails.

\section{Possible Generalizations}
\label{sec:PG}

Let us remind the reader that the above results on $r_d$-dependence are general
while those on $L$-dependence are formally obtained only when the respective
anomalous exponent $\Delta$ is much less than the normal exponent $2\gamma$.
The whole approach of Sect.2 was developed
for the fourth-order correlation function at $d>2$.
The generalization for the $n$-th correlation function at $d> n-2$ is
straightforward by means of the same representation of $\hat{\cal L}$ in
terms of $r_{ij}$. Such a generalization leads to the
new predictions for $n>4$: the scaling exponent $2\gamma-\Delta$
should appear at all high-order functions. The direct check shows that
the respective zero modes appear in the partially reducible contributions only.
For example, the six-point object
$F_{123456}\equiv \langle\overline{
\theta({\bf r}_1)\theta({\bf r}_2)\theta({\bf r}_3)\theta({\bf r}_4)
\theta({\bf r}_5)\theta({\bf r}_6)}\rangle$ should necessarily contain
terms like $f(r_{12})\Gamma_{3456}$ i.e. $r^{2\gamma-\Delta}L^{\gamma+\Delta}$.
Such contributions should cancel out in structure functions like
$\langle\langle
(\theta_1-\theta_2)^{n}
\rangle\rangle$. To find the scaling exponents of the $n$-th structure
functions
in the limit $d\rightarrow\infty$,
one should iterate the zero mode with the scaling exponent $n\gamma$, which
will be published elsewhere.

Quite a different picture may appear for $d\leq n-2$. In this case, $n(n-1)/2$
distances between points are not longer the independent variables. Additional
constraints should be imposed on that set of (otherwise very convenient)
variables, which may lead to the possibility of additional zero modes.

As a final remark, let us emphasize that the above results are valid for a
Gaussian
pumping only and discuss what might be the consequences of the pumping
non-Gaussianity. Let us add to the fourth-order pumping correlation function
(\ref{xi4}) the irreducible part $
\chi_4({\bf r}_1,{\bf r}_2,{\bf r}_3,{\bf r}_4)
\delta(t_1-t_2)\delta(t_1-t_3)\delta(t_1-t_4)$.
One may model the function $\chi_4$ by the step function: $
\chi_4^m({\bf r}_1,{\bf r}_2,{\bf r}_3,{\bf r}_4)=
P_4$ if all $r_{ij}<L$ and zero if any $r_{ij}>L$.
The production rate of $\theta^4$ is $3P_2f(0)+P_4$ (let us remind that
the advection preserves the integral of any power of the scalar field).
The ratio $\tau_*=P_4/P_2^2$ having the dimensionality
of time is a correlation time of the pumping, which is assumed to be the
smallest
time in the problem. As a result, the terms generated by $\chi_4$ will be small
yet
we keep them to show what qualitatively new may appear in the case of a
non-Gaussian pumping with a finite correlation time.
The equation (\ref{b7a}) acquires additional term:
$$-\hat{\cal L}\Gamma_{1234}=\chi_4+\Phi_{12;34}+\Phi_{13;24}+\Phi_{14;23}.$$
The presence of the function $\chi_4$ changes scaling in the
convective interval, it causes the terms
with anomalous scaling $\tau_*r_{ij}^\gamma$ to appear in
$\Gamma$ which corresponds to the scaling $\zeta_n=\zeta_2$
discussed in \cite{94Fal-a,94Kra}. Also, the terms with $r_{12}^0$
appear in $\langle\langle
\epsilon_1\theta_2^2\rangle\rangle$, while no additional terms appear in
$\langle\langle\epsilon_1\epsilon_2\rangle\rangle$.
Note that within our approach the $\tau_*$-related terms are small corrections
even at the diffusion scale despite the fact that
they decrease slower than the main terms as distances decrease.
Most probably, finite correlation time of the pumping will
lead to a substantial contributions with an another
anomalous scaling yet this is beyond
the scope of the present analysis. The development
of the theory for the finite correlation times of the velocity and pumping is
necessary for a meaningful comparison between theory and experiment. This will
be the subject of future publications.
On the other hand, the more detailed experimental
results are desirable, in particular, direct measurements of
the correlation function of the dissipation field.

\section{Conclusion}
\label{sec:disc}

We have shown that the fourth-order correlation function $F_{1234}$
has the scaling exponent $2\gamma-\Delta$
in the convective interval at $d>2$. The anomalous exponent
$\Delta$ is analytically found at $d\rightarrow\infty$. Considering the
behavior of
$F_{1234}$ in the convective interval where some separations $\rho$ tend
to zero we established that the $\rho$-dependent contribution to $F_{1234}$
also tends to zero. Moreover, nothing terrible happens when $\rho$ passes
through the diffusion scale $r_d$ and we find that
\begin{equation}
\langle (\theta_1-\theta_2)^{4}\rangle
\propto |{\bf r}_1-{\bf r}_2|^{2\gamma-\Delta}L^\Delta,
\label{cs0} \end{equation}
for $r_{12}$ from the convective interval.

We analyzed also  the case where two separations are much smaller
than the distance $R$ between the pairs and found the  contribution,
mixing both small distances, which determines the correlator
\begin{eqnarray} &&
\langle(\theta_1-\theta_2)^2
(\theta_3-\theta_4)^2\rangle
-\langle(\theta_1-\theta_2)^2\rangle
\langle(\theta_3-\theta_4)^2\rangle\,,
\label{cs} \end{eqnarray}
at $r_{12},r_{34}\ll R$.
We established that the scaling behavior of that correlation function in
the convective interval is characterized by the additional anomalous
exponent $\delta$ defined by (\ref{delta}). The corresponding contributions to
the irreducible part $\Gamma_{1234}$ of the correlator $F_{1234}$ are
presented in (\ref{ZM}). The isotropic contribution (\ref{ZMa})
behaves $\propto r_{12}^\gamma r_{34}^\gamma (L/R)^\Delta$, while the
anisotropic
one (\ref{ZMc}) behaves $\propto r_{12}^\delta r_{34}^\delta
R^{(2\gamma-2\delta)}(L/R)^\Delta$. The cross term (\ref{ZMb-c}) behaves
$\propto (r_{12}^\delta r_{34}^\gamma +r_{34}^\delta r_{12}^\gamma)
R^{\gamma-\delta}(L/R)^\Delta $. The analysis of the continuation of the
correlation
function (\ref{cs}) for $r_{12}$, $r_{34}$ passing to the diffusive
interval gives the following scaling laws
\begin{eqnarray} &&
\langle\langle\epsilon_1\epsilon_3\rangle\rangle=\kappa^2
\langle\langle(\nabla\theta_1)^2(\nabla\theta_3)^2\rangle\rangle
\propto r_d^{0} (L/r_{13})^\Delta \,,
\label{cs1} \\ &&
\langle\langle(\nabla_\alpha\theta_1\nabla_\beta\theta_1-
d^{-1}\delta_{\alpha\beta}(\nabla\theta_1)^2)
\nonumber \\ &&\times
(\nabla_\mu\theta_3\nabla_\nu\theta_3-
d^{-1}\delta_{\mu\nu}(\nabla\theta_3)^2)\rangle\rangle
\propto r_d^{2\delta -4} r^{2\gamma-2\delta}_{13}(L/r_{13})^\Delta\,,
\nonumber\\&&\label{cs2} \\ &&
\langle\langle\epsilon_1[\nabla_\alpha\theta_3\nabla_\beta\theta_3
-d^{-1}\delta_{\alpha\beta}(\nabla\theta_3)^2]\rangle\rangle
\propto r_d^{\delta-2} r^{\gamma-\delta}_{13}(L/r_{13})^\Delta\,,
\nonumber\\&&\label{cs9} \end{eqnarray}
where the separation $r_{13}$ lies in the convective interval.
The double angular brackets designate irreducible correlation functions.
The analogous analysis can be performed for the correlation function
\begin{equation}
\langle(\theta_1-\theta_2)^2\theta^2_3\rangle
-\langle(\theta_1-\theta_2)^2\rangle\langle\theta^2_3\rangle
\label{nn} \end{equation}
at $r_{12}\ll r_{13}$. The contribution (\ref{g11}) gives the law
$\propto r_{13}^\gamma r_{12}^\gamma$ for the isotropic part
of the correlation function (\ref{nn}) what leads to
\begin{equation}
\langle\langle\epsilon_1\theta_3^2\rangle\rangle \propto
r_d^{0}r_{13}^\gamma(L/r_{13})^\Delta \,.
\label{cs3} \end{equation}
The anisotropic part of the correlation function (\ref{nn})
is determined by the contribution (\ref{ZMb}) what is proportional to
$r_{13}^{2\gamma-\delta}r_{12}^\delta$, it leads to
\begin{equation}
\langle\langle\theta_1^2[\nabla_\alpha\theta_2\nabla_\beta\theta_2
-d^{-1}\delta_{\alpha\beta}(\nabla\theta_2)^2]\rangle\rangle
\propto r_d^{\delta-2} r_{13}^{2\gamma-\delta} (L/r_{13})^\Delta\,.
\label{cs8} \end{equation}

The appearance of $r_d$-dependence in the correlation functions
(\ref{cs1}--\ref{cs8}) may correspond to the presence of the ultraviolet
divergences found in \cite{94LL,94LPF} in the diagrams for powers of
gradients both of the velocity and of the passive scalar.
That means that, contrary to $L$-related scaling, $r_d$-related anomalous
scaling of the scalar derivatives could be caught perturbatively. Moreover,
in the asymptotic limits of the normal overall scaling of $\Gamma_{1234}$
(at $d\to\infty$ or $\gamma\to 2^-$)
it is tempting to describe the structure of
(\ref{cs0},\ref{cs1},\ref{cs2},\ref{cs9},\ref{cs3},\ref{cs8})
using the language of the so-called operator algebra \cite{69Pol,69Kad,69Wil}
(developed in the context of second-order phase transition theory)
the validity of which for turbulence was argued in \cite{94LL}. Namely,
at $d\to\infty$ or $\gamma\to 2^-$ in
accordance with those expressions, the passive scalar $\theta$ has the
dimensionality $-\gamma/2$, $(\nabla\theta)^2$ has the dimensionality $0$ and
$(\nabla_\alpha\theta\nabla_\beta\theta-
d^{-1}\delta_{\alpha\beta}(\nabla\theta)^2)$
has the dimensionality $\delta-\gamma$. In the general case of arbitrary $d$
and
$\gamma$ when
the $L$-related anomalous scaling does exist, the underlying algebraic
structure (if it exists at all) is unclear at the moment.

All said above concerns the terms associated with zero modes
of the operator ${\cal L}$ in the convective interval. There exist
also a forced solution in the convective interval originating from
(\ref{per3}). The tails in the diffusion intervals created by this
forced solution can be interpreted as zero modes (in the diffusive
interval only) which are determined by the estimates (\ref{ZN}).
Those zero modes produce the contributions to the correlation
functions (\ref{cs2},\ref{cs9},\ref{cs8}) which are
$\propto r_{13}^{-\gamma}(L/r_{13}^\gamma)$ and prevail if $\delta>3\gamma/2$,
$\delta>\gamma/2$ and $\delta>3\gamma$ correspondingly.
Those contributions are not governed by the operator algebra
even in the asymptotic cases of $d\to\infty$ or $\gamma\to 2^-$.
It is natural from the diagrammatic point of view since those contributions
correspond to ``one-bridge'' diagrams which does not contain series
producing anomalous scaling. Thus the factor $\propto r_{13}^{-\gamma}$
is determined simply by the ``bridge'' factor
$\nabla\nabla{\cal K}\propto R^{-\gamma}$. Of course analogous
contributions exist in all above correlation functions but
only in (\ref{cs2},\ref{cs9},\ref{cs8}) they can prevail at some $\delta$.

To conclude, $1/d$-expansion tells us that the anomalous scaling is present
already in the oversimplified model of the delta-correlated velocity field.
If one formally uses our formula (\ref{Delta}) for the anomalous exponent
at Richardson-Kolmogorov case $d=3,\gamma=2/3$ it gives $\Delta=16/9$ which
substantially larger than the experiments and numerics give
($\Delta\simeq0.4\div
0.5$), the difference could be accounted by inaccuracy of both an asymptotic
formula at finite $d$ and the model with delta-correlated velocity.
Note that our results are valid not only for steady state but also for
decaying turbulence of a passive scalar with a compact spectrum
as an initial condition. In that case, the scalar cascade is accelerated
as it goes towards small scales (with typical time $t\propto r^{(2-\gamma)/2}$)
so that the small-scale part of the scalar distribution is quasi-steady.

\acknowledgements

We are indebted to E. Balkovsky, V. Belinicher, A. Fairhall, U. Frisch, O. Gat,
V. L'vov, A. Migdal, M. Nelkin, I. Procaccia and V. Yakhot for discussing
the manuscript of the present paper. Two of us (G.F. and V.L.)
are grateful to I. Procaccia for organizing the excellent
Eilat workshop where some of the above results have been extensively discussed.
We are deeply grateful to R. Kraichnan for numerous illuminating explanations.
This work was partly supported by the Clore Foundation (M.C.)
by the Rashi Foundation (G.F.)
and by the Minerva Center for Nonlinear Physics (I.K. and V.L. ).

\end{multicols}
\end{document}